\documentclass[aps,prl,preprint,superscriptaddress]{revtex4-1}

\usepackage{graphicx}
\usepackage{amsmath}
\usepackage{mathrsfs}
\usepackage{amssymb}
\usepackage{url}
\usepackage{graphicx}
\usepackage{color}
\usepackage{soul}
\usepackage{hyperref}
\usepackage{mathtools}
\usepackage[detect-none]{siunitx}
\usepackage{wasysym}
\usepackage[dvipsnames]{xcolor}

\begin{document} 

\title{Optical vortex crystals with dynamic topologies}

\author{Marco Piccardo}
\email[]{marco.piccardo@iit.it}
\affiliation{Center for Nano Science and Technology, Fondazione Istituto Italiano di Tecnologia, Milan, Italy}

\author{Michael de Oliveira}
\affiliation{Center for Nano Science and Technology, Fondazione Istituto Italiano di Tecnologia, Milan, Italy}
\affiliation{Physics Department, Politecnico di Milano, Milan, Italy}

\author{Andrea Toma}
\affiliation{Fondazione Istituto Italiano di Tecnologia, Genoa, Italy}

\author{Vincenzo Aglieri}
\affiliation{Fondazione Istituto Italiano di Tecnologia, Genoa, Italy}

\author{Andrew Forbes}
\affiliation{School of Physics, University of the Witwatersrand, South Africa}

\author{Antonio Ambrosio}
\email[]{antonio.ambrosio@iit.it}
\affiliation{Center for Nano Science and Technology, Fondazione Istituto Italiano di Tecnologia, Milan, Italy}

\maketitle

\textbf{Vortex crystals are geometric arrays of vortices found in various physics fields, owing their regular internal structure to mutual interactions within a spatially confined system. In optics, vortex crystals may form spontaneously within a nonlinear resonator but their usefulness is limited by the lack of control over their topology. On the other hand, programmable devices used in free space, like spatial light modulators, allow the design of nearly arbitrary vortex distributions but without any intrinsic dynamics. By combining non-Hermitian optics with on-demand topological transformations enabled by metasurfaces, we report a solid-state laser that generates vortex crystals with mutual interactions and actively-tunable topologies. We demonstrate  10$\times$10 coherent vortex arrays with nonlocal coupling networks that are not limited to nearest-neighbor coupling but rather dictated by the crystal's topology. The vortex crystals exhibit sharp Bragg diffraction peaks, witnessing their coherence and high topological charge purity, which we resolve spatially over the whole lattice by introducing a parallelized analysis technique. By structuring light at the source, we enable complex transformations that allow to arbitrarily partition the orbital angular momentum inside the cavity and to heal topological charge defects, making these resonators a robust and versatile tool for advanced applications in topological optics.}

\medskip

In many natural systems, ensembles of vortices can self-organize into regular patterns through their mutual interaction \cite{Jin2000}. This is the case for
%quantum vortices in a superfluid $^4$He nanodroplet \cite{Gomez2014},
clusters of cyclones encircling the poles of Jupiter \cite{Adriani2018}, knotted lattices in chiral liquid crystals \cite{Tai2019}, and motile haploid cells of sea urchins swimming in vortex trajectories \cite{Riedel2005}. In broad-area lasers, crystal-like arrangements of optical vortices can also form spontaneously, due to the nonlinear interaction of several competing transverse modes \cite{Brambilla1991,Scheuer1999}. These crystals consist of arrays of intensity nulls, each corresponding to a phase singularity of the field carrying orbital angular momentum (OAM) with a topological charge \cite{Coullet1989,Allen1992}. The self-organizing nature of these vortex crystal lasers, while fascinating as a nonlinear dynamic system, makes them unsuitable for practical applications, as the laser states can be nonstationary \cite{staliunas2003transverse} and their topology cannot be tuned at will. Thus, the applications requiring vortex crystals \cite{Wang2018,Shen2019}, including parallelized super-resolution imaging \cite{Vicidomini2018},
%object spinning detection in optical metrology [Lavery2013],
OAM-multiplexed communications \cite{Ren2016}, and multiparticle micromanipulation with optical tweezers \cite{Padgett2011,Woerdemann2013},
%and 3D displays \cite{Li2016},
have turned their attention to alternative generation schemes, ranging from spatial light modulators (SLMs) and Dammam gratings \cite{Lei2015} to metamaterials \cite{Kildishev2013,Yu2014,Maguid2018,Jin2017,Piccardo2020APL}. However, all of these extracavity implementations carve out the vortices from an input beam profile, resulting in reduced efficiency. Moreover, these devices only realize the photonic design programmed by the user without exhibiting an internal dynamic evolution, thus losing the appeal of an active system. A recent approach is to coherently combine multiple microring lasers based on supersymmetry, which is promising for power scaling but is currently limited to a topological charge of one \cite{Qiao2021}.

In this work, we present a structured light laser \cite{Forbes2019} capable of producing vortex crystals with a topology that is tailored directly at the source. Just as in a broad-area laser, our gain medium has a large transverse cross section supporting many transverse modes. Instead of allowing these modes to self-organize in the resonator, we force them to form a lattice of a hundred laser beams by inserting a mask that modulates the phase and amplitude of the field \cite{Nixon2013,Pal2017}. The phase profile is designed to impart arbitrary OAM values to each beam of the lattice and in doing so, we artificially introduce an amount of OAM in the system that must be conserved. This effectively partitions the resonator into two sections---one on either side of the OAM-transforming mask--- each emitting a laser lattice imbued with a different topological charge. We spatially couple the vortices using a non-Hermitian mechanism that introduces dissipative losses in the resonator, making the system stable and coherent. The coupling network is not limited to nearest-neighbour interactions as for arrays of Gaussian lasers with no topological charge, but instead can be tailored to mix vortices that are widely separated in the lattice. While different schemes for generating a single vortex inside a resonator have been demonstrated \cite{Naidoo2016,Maguid2018,Sroor2020,Ito:10,Cai2012,Huang2020}, this work accomplishes the conditions for generating vortex crystals in a single laser cavity, realizing a platform to explore complex topological transformations and collective vortex effects at the source.

\medskip
\textbf{\large{Results}}

\textbf{Operating principle of the metasurface laser}

\begin{figure}[h!]
    \centering
    \includegraphics[width=0.67\textwidth]{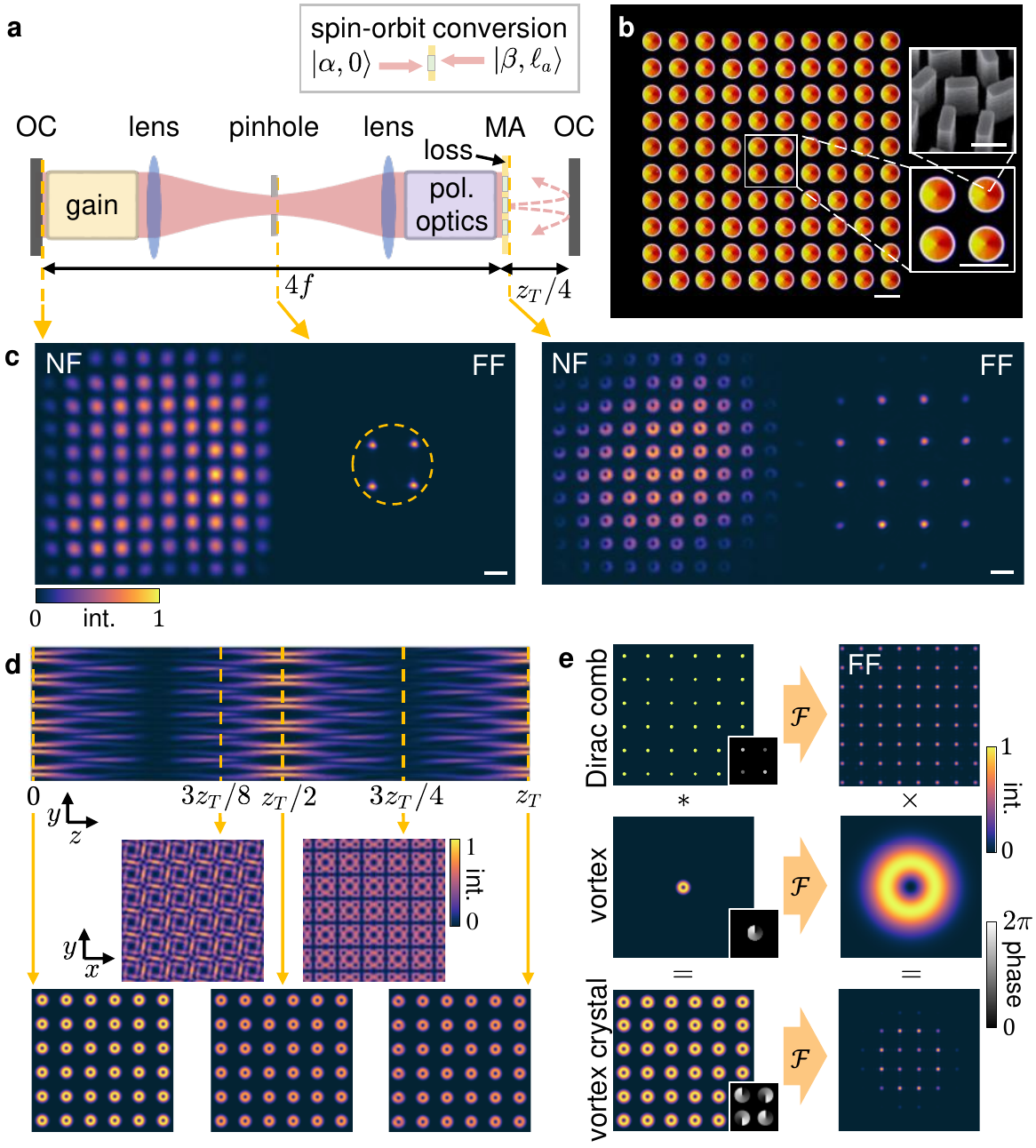}
    \caption{}
    \label{fig_MSlaser}
\end{figure}

The resonator comprises of a diode-pumped neodymium-doped yttrium aluminum garnet (Nd:YAG) laser with high Fresnel number in a self-imaging ($4f$) configuration (Fig. \ref{fig_MSlaser}a) supporting a very large number of transverse modes ($\sim10^5$) \cite{Cao2019}. Two intracavity lenses, which form a telescope system, image the plane of the metasurface array onto the opposite end of the cavity, where an output coupler (OC) is present (Materials and Methods). The other OC is displaced from the metasurface array to couple the laser beams via the Talbot effect. The metasurface array provides the necessary intracavity topological charge conversion. It consists of 10$\times$10 metasurfaces in a square lattice geometry with amorphous Si nanopillars on a fused SiO$_2$ substrate (Fig. \ref{fig_MSlaser}b). Different vortex-plate designs are possible, either based on the Pancharatnam-Berry phase, such as $q$-plates \cite{Marrucci2006,Devlin2017Q} that are used in this work, or on the propagation phase, such as $J$-plates \cite{Devlin2017J} (Supplementary Materials). The general requirement is that the circulating beam does not acquire any net topological charge upon a double pass through the metasurface array. This condition cannot be satisfied using a polarization-independent spiral phase plate in a plane-parallel resonator \cite{Wen2021}, but we achieve it by exploiting the engineered birefringence of a metasurface, in combination with the appropriate intracavity polarization optics, for complete spin-orbit control of the beam (for details on how the roundtrip condition is satisfied, see the Supplementary Materials).

The telescope configuration gives direct experimental access to the far-field (FF) plane, which is physically located between the two intracavity lenses and contains the optical Fourier transform (FT) of the near-field (NF) of the metasurface array. By placing a pinhole in the FF it is possible to filter high-spatial frequency components of the beam without reducing the gain effective area nor affecting the laser efficiency \cite{Nixon2013a,Cao2019,Tradonsky2017}. Since the metasurface array separates the cavity into two sections with different topological charges, a FF pinhole allows us to force the existence of the Gaussian beams---having the smallest spatial frequency content in the FF---on the telescope segment of the cavity, leaving the vortex beams in the Talbot section. 

The interstices between the metasurfaces of the array are filled with a metallic mask that spatially filters the beams' transmission. The beams diffract at every metasurface aperture, resulting in an array of interfering vortices (Supplementary Movies 1,2). Due to the periodic arrangement of the vortices, the Talbot effect ensures that self-images of the array are created at specific planes along the propagation direction, as shown in the Talbot carpet of Fig. \ref{fig_MSlaser}d, whose positions in general depend on the phase relation among the vortices. At the characteristic Talbot distance $z_T = 2d^2/\lambda$ of a square array, where $d = 0.3$ mm is our period and $\lambda = 1064$ nm is the laser wavelength, both an array of in-phase and out-of-phase vortices---where the nearest neighbors have a phase difference of 0 and $\pi$, respectively---will form a self-image \cite{Tradonsky2017}. This degeneracy of the phase solutions is lifted at $z_T/2$, where only an out-of-phase array can produce a self-image, so we use this distance in our system. Since the Talbot propagation path is folded in the resonator, this requires us to set a distance of $z_T/4 = 42$ mm between the metasurface array and the nearest OC (Fig. \ref{fig_MSlaser}a).

The generation of a vortex crystal is demonstrated in Fig. \ref{fig_MSlaser}c. Different near-field intensity distributions are observed at each end of the cavity: one corresponding to an array of Gaussian beams, the other to an array of donut beams with dark cores, signaling the occurrence of a topological charge conversion inside the cavity. The phase coherence of the arrays is witnessed by the Bragg diffraction peaks present in the far-field distribution, obtained by a FT of the near-field distributions. The Bragg pattern of the vortex crystal can be easily understood using the convolution theorem. Since a vortex crystal corresponds to the convolution of a Dirac comb with a vortex beam (first column of Fig. \ref{fig_MSlaser}e), its FT is simply given by the pointwise product of the FTs of the Dirac comb and the vortex, resulting in a single large pixelated vortex (second column of Fig. \ref{fig_MSlaser}e). The experimental results are in very good agreement with numerical Fox-Li simulations of the laser, which include the saturable gain nonlinearity, proving that the measured lasing mode is the one experiencing the lowest loss in the cavity (Fig. S6, Materials and Methods).

\medskip
\textbf{Parallelized topological charge analysis}

\begin{figure}[h!]
    \centering
    \includegraphics[width=1\textwidth]{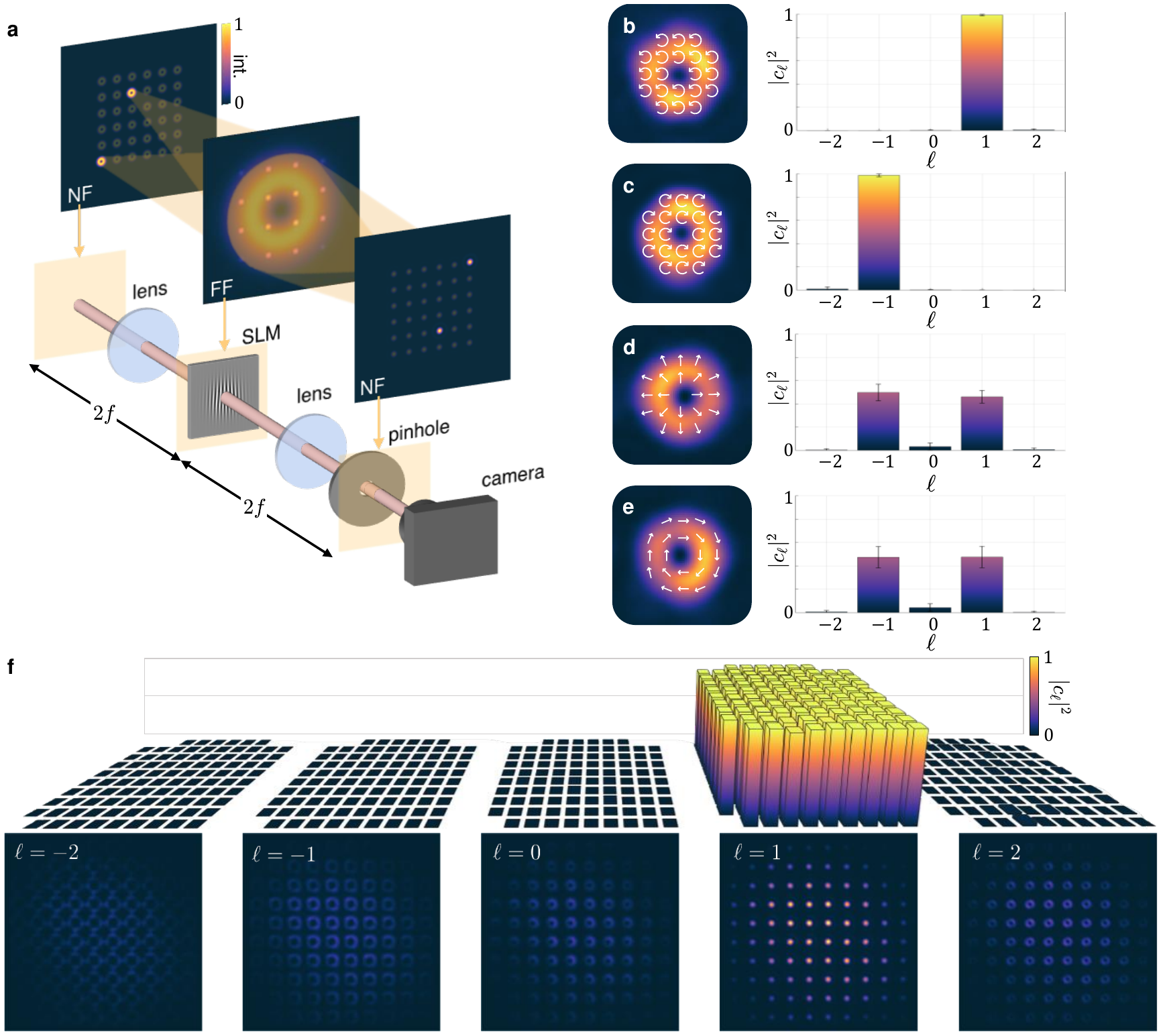}
    \caption{}
    \label{fig_chargeanalysis}
\end{figure}

The Bragg diffraction peaks are a signature of phase-locking in the vortex crystal but they do not directly show the value of the topological charge carried by the vortices. A standard approach to analyze the topological charge of a vortex is to perform a modal decomposition by measuring the inner-product between the field and appropriate correlation filters. In general, a single beam could be selected by applying a digital aperture on an SLM in the near-field of the array and its charge revealed by displaying a phase and amplitude hologram of the opposite charge, leading to an intensity peak at the beam's optical axis in the far-field \cite{Pinnell2020}. However, this is an inefficient technique for an array, with a long sequential processing time proportional to the number of vortices in the crystal. By exploiting the fact that each vortex in the crystal is imaged to the same far-field profile regardless of its position in the array (Fig. \ref{fig_chargeanalysis}a), and by using the linearity of the FT,
%(i.e. the FT of a sum of functions is the sum of the FTs of the functions)
we introduce a parallelized technique to analyze the topological charge of each beam in the array all at once. Using an SLM we carry out a modal decomposition of all the vortices in the far-field and measure their respective on-axis intensity in the near-field (Materials and Methods). The modulated beams will now have a charge given by the sum of the decomposition charge displayed on the SLM and the intrinsic one of the crystal. This corresponds to an array of Gaussians when the SLM and crystal charges have equal values but opposite signs, and to an array of donuts of variable radii in all other cases (Fig. \ref{fig_chargeanalysis}f). In Fig. \ref{fig_chargeanalysis}b--e we present the results for a  metasurface array with charge $\ell_a = 1$. The measured charge purity statistically averaged over the array is near 100$\%$ for right- and left-circularly-polarized light. By changing the intracavity polarization optics it is also possible to generate vector vortex crystals with azimuthally and radially polarized light, corresponding to a superposition of opposite charges ($\ell = \pm1$, Fig. \ref{fig_chargeanalysis}d,e). In this case a small contribution at $\ell=0$ is also present (4$\%$), due to a residue of the beam that is not converted by the metasurface array and not filtered outside the cavity (Supplementary Materials). In the rest of this work we will concentrate on scalar vortex beams, keeping the focus on the OAM properties of the laser.

\medskip
\textbf{Scaling up the topological charge}

\begin{figure}[h!]
    \centering
    \includegraphics[width=1\textwidth]{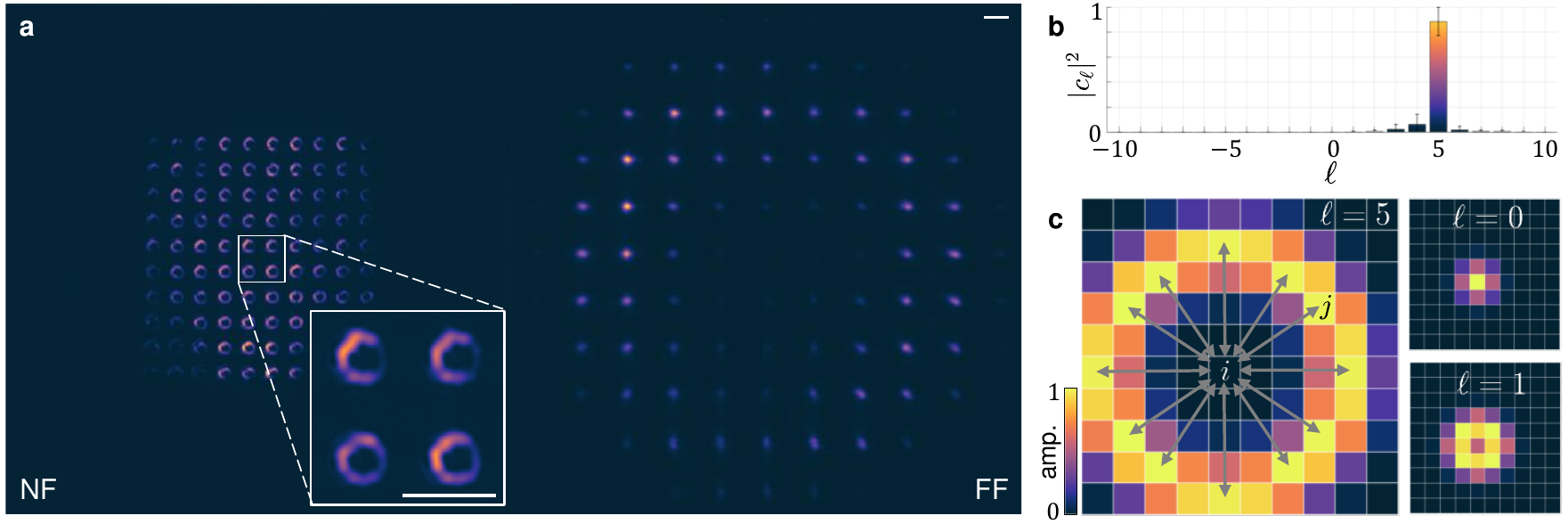}
    \caption{\textbf{Increasing the vortex crystal charge and coupling networks.} (\textbf{a}) Experimental near- and far-field distributions of a vortex crystal with $\ell=5$, showing large dark cores. All scale bars are 300 $\mu$m. (\textbf{b}) Topological charge spectrum of the vortices averaged over the whole crystal.
    %The spin-orbit scheme used here produces beams with uniform circular polarization.
    (\textbf{c}) Magnitude of the coupling coefficient $|\kappa_{i,j}|$ calculated between each vortex in the array (indexed by $j$, with $j$ varying over all array positions) and the one located at the center (indexed by $i$). The size of the coupling network increases with $\ell$. The dominant contributions in the $\ell = 5$ case are marked by arrows. In all plots the grids define the positions of the array.}
    \label{fig_L5}
\end{figure}

Our metasurface laser scheme allows to scale up the charge of the vortex crystal by increasing the azimuthal gradient in the phase design of the metasurfaces. We characterized an array with a charge of 5 (Fig. S7). Due to a charge-dependent increase in vortex core, the typical diameter of the donuts observed in the near-field is $150\pm10$ $\mu$m (Fig. \ref{fig_L5}a), larger than that of the charge-1 donuts shown above with $91\pm5$ $\mu$m (Fig. \ref{fig_MSlaser}c)---a size change in agreement with the numerical simulations ($138\pm8$ $\mu$m and $90\pm9$ $\mu$m, respectively). The larger topological charge content of the vortex crystal is also manifested in the far-field distribution (Fig. \ref{fig_L5}a), corresponding again to a pixelated vortex but with a dark core larger than the charge-1 case.

The topological charge purity remains high, with a peak at $\ell=5$ around 90$\%$ (Fig. \ref{fig_L5}b). The small spurious contributions at other charges are due to imperfect Talbot reconstruction in the peripheral regions of the array. This can be understood by examining the coupling network of the array, which is more complicated than for crystals without topological charge, where coupling is dominated by nearest-neighbor interactions ($\ell=0$ in Fig. \ref{fig_L5}c). The coupling matrix $\kappa_{i,j}$ of the vortex crystal is non-Hermitian \cite{Longhi2018,arwas2021anyonic}, being complex, with an imaginary component due to transmission losses at the mask, and symmetric, like the field propagation from one vortex to another, thus $\kappa^\ast_{i,j} \neq \kappa_{j,i} = \kappa_{i,j}$. We calculate the magnitude of $\kappa_{i,j}$ for any position $j$ in the array and $i$ corresponding to a central position \cite{Tradonsky2017}. Considering a vortex crystal with $\ell=5$ (Fig. \ref{fig_L5}c), the main power coupling comes from relatively distant vortices lying on a large circle. When $i$ corresponds to the position of a vortex far from the center of the array the power coupling will be anisotropic, due to the finite size of the array, forming a vortex with an asymmetric donut intensity profile and fractional OAM producing small sidebands in the topological charge spectrum. These effects are less pronounced for small charges (Fig. \ref{fig_L5}c, $\ell=1$) and can be mitigated by increasing the array size.

%since this increases the quality of the self-imaging reconstruction, as shown by the higher transmission through the Talbot section reduces spillover  (Fig. \ref{fig_L5}E).
%A higher transmission results in ahigher-quality self-imaging reconstruction and a higher topological charge purity, as averaged over the array.

\medskip
\textbf{Partitioning orbital angular momentum inside the cavity}

The spin-orbit transformations implemented in our cavity in principle allow for several topological solutions, each comprising a pair of vortex crystals with different topological charge occupying the two sides of the metasurface array. We denote a generic topological solution as $(m~|~n)$, where $m$ and $n$ are the charges of the crystals on the telescope and Talbot segments of the cavity, respectively. OAM conservation imposes the condition $m + \ell_a = n$, where $\ell_a$ is the metasurface array charge. Until here, the use of the far-field pinhole forced the appearance of beams with zero charge in the telescope section, imposing the solution $(0~|~\ell_a)$.

\begin{figure}[h!]
    \centering
    \includegraphics[width=1\textwidth]{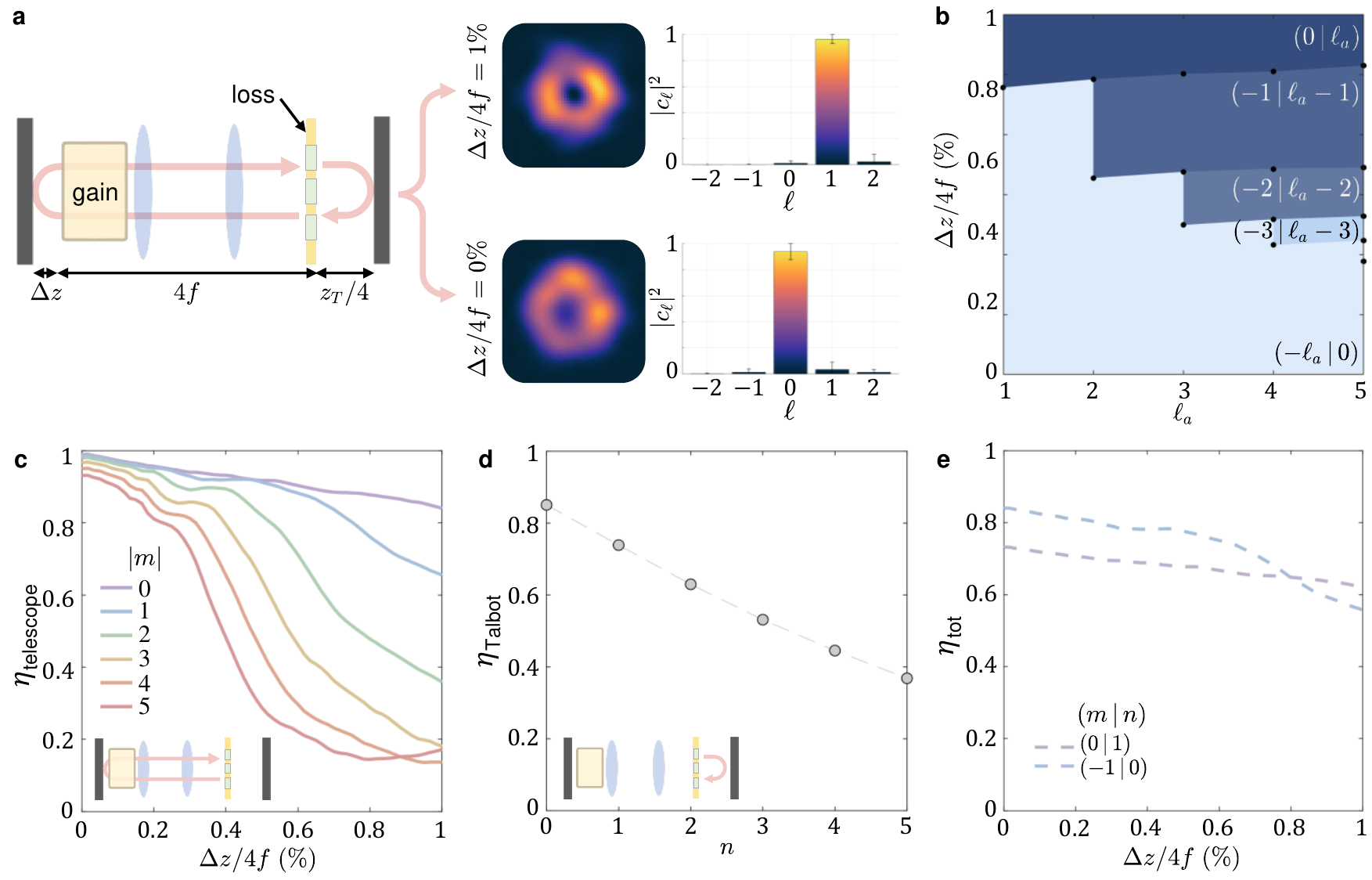}
    \caption{}
    \label{fig_degeneracylift}
\end{figure}

After removing the far-field pinhole, there could be many degenerate solutions. However, this degeneracy is lifted by two main elements in the cavity giving OAM-dependent losses and favoring a specific solution.
%For instance, in a telescopic cavity with a single OAM-converter it was previously found that the vortex prefers the side of the resonator with the active medium---when this is placed in the FF region---due to its larger modal volume as compared to a Gaussian beam and increased overlap with the gain medium \cite{Maguid2018}. In our resonator the active medium lies in the NF region, next to one of the OCs, where the beams are collimated and the modal volume plays a minor role.
Both types of losses are dissipative and originate from the spatial filtering of the mask, after the beam has made a roundtrip in each segment of the cavity (Fig. \ref{fig_degeneracylift}a). The first contribution relates to a roundtrip in the telescope section. We find that a small deviation $\Delta z$ in cavity length from the ideal $4f$ value produces an OAM-dependent transmission efficiency through the mask, due to the divergence of the beams (Fig. \ref{fig_degeneracylift}c). The second contribution is connected with a roundtrip in the Talbot section. Due to the finite size of the array, part of the Talbot self-image does not overlap with the original array but rather spills out around its edges \cite{Mehuys:91}, getting blocked by the mask. Since the spillover increases with the divergence of the diffracting beams, the transmission efficiency diminishes with the charge (Fig. \ref{fig_degeneracylift}d). The two described effects are in competition and the lowest-loss solution chosen by the laser depends on $\Delta z$.

By combining the telescope and Talbot efficiencies, we calculate the total transmission efficiency for a full roundtrip in the resonator. As shown in Fig. \ref{fig_degeneracylift}e, in the case of $\ell_a = 1$ there are two possible topological solutions, $(0~|~1)$ and $(-1~|~0)$, dominating in different ranges of the tuning parameter with a transition point around $\Delta z/4f = 0.8\%$. We verify experimentally the existence of these two solutions by displacing the OC on the telescope side by a few millimeters and measuring the change in topological charge spectrum of the crystal emitted from the Talbot side of the laser (Fig. \ref{fig_degeneracylift}a). Remarkably, the beams are donut-shaped also when they do not carry any charge. This feature arises from the phase-only transformations operated by the metasurfaces, which unwind the phase but leave an amplitude term in the near-field of the beams, as also observed in the laser simulations (Supplementary Materials). These findings can be generalized for higher array charges resulting in a set of $\ell_a+1$ different topological solutions, ranging from $(0~|~\ell_a)$ to $(-\ell_a~|~0)$ (Fig. \ref{fig_degeneracylift}b).
%The results obtained with this model agree with the output of numerical Fox-Li simulations, though some small fluctuations in the values of critical $\Delta z$ separating the different solutions may occur, due to the active beam waist adjustment occurring in the laser over many iterations.

\medskip
\textbf{Healing of defects}

%cavity deltaz is tuned to impose a 0|\ell solution without FF pinhole

\begin{figure}[h!]
    \centering
    \includegraphics[width=0.67\textwidth]{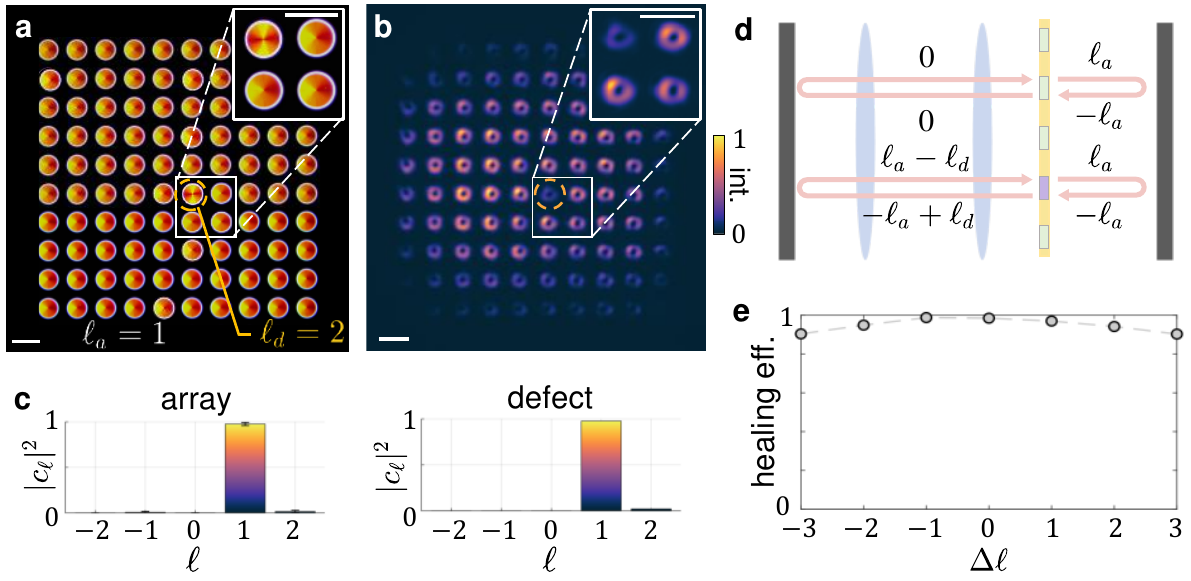}
    \caption{\textbf{Healing of a topological charge defect.} (\textbf{a}) Optical microscope image of a metasurface array with charge $\ell_a = 1$ everywhere except for an intentional defect with charge $\ell_d = 2$ (encircled device). (\textbf{b}) A vortex crystal emitted from the metasurface laser showing an array of donuts with similar size. All scale bars are 300 $\mu$m. (\textbf{c}) Topological charge spectra measured at the defect position (right) and averaged over the rest of the array (left), demonstrating the healing of the charge defect. (\textbf{d}) Topological charge evolution in a cavity roundtrip for a double pass through the defect device (bottom) or any other metasurface in the array (top). (\textbf{e}) Simulated defect healing efficiency, defined as the fraction of beam power at the defect location carrying a charge equal to that of the vortex crystal. The efficiency is shown as a function of the charge difference between the defect and the array $\Delta \ell = \ell_d - \ell_a$, where $\ell_a = 1$ as in the experiment.}
    \label{fig_defect}
\end{figure}

So far, we have presented a vortex crystal topology determined by arrays of identical metasurfaces. Next, we break the symmetry of the array by introducing an intentional topological charge defect and study the properties of the resulting vortex crystal. We fabricate a $10\times10$ array with metasurfaces carrying a charge $\ell_a = 1$ except for a defect device with charge $\ell_d = 2$ located in the central region of the array (Fig. \ref{fig_defect}a). We remove the FF pinhole from the cavity and tune its length to sustain the $(0~|~\ell_a)$ solution. At the output of the metasurface laser, we observe a vortex crystal with donut beams of the same size (Fig. \ref{fig_defect}b), as in the defect-free laser (Fig. \ref{fig_MSlaser}c). Further examination of the spatially-resolved topological charge spectrum reveals that in correspondence to the defect metasurface the dominant OAM component is $\ell=1$, as for the rest of the crystal, indicating the ability of the metasurface laser to heal the topological charge defect (Fig. \ref{fig_defect}c).

The robustness to defects is reminiscent of the property of topological insulator lasers \cite{Bandres2018}, although it is achieved here without resorting to topological protection---a concept that does not easily lend itself to solid-state lasers---but rather by exploiting the non-Hermitian nature \cite{Longhi2018, ElGanainy2019} of the vortex coupling mechanism. The defect recovery engine built into our laser is based on the OAM version of the Talbot self-healing effect \cite{Dammann1971}: multiple passes through the Talbot section of the cavity remove non-periodic topological faults in the array, strengthening only the periodic components at the self-imaging plane. Following a full roundtrip through the cavity, we can see how the topological charge is repeatedly converted at the defect metasurface, resulting in a self-consistent solution (Fig. \ref{fig_defect}d; for the defect healing transient, see Supplementary Movie 3). The additional charge introduced by the defect device does not disappear from the system but instead manifests itself in the telescope section as a beam with a new charge $\pm(\ell_a - \ell_d)$. As long as this beam can propagate through the telescope segment, any value of the topological charge of the defect will be recovered in the vortex crystal with high efficiency (Fig. \ref{fig_defect}e). Since this beam carries a charge larger than the surrounding array of Gaussian beams (i.e., in our experiment $|\ell_a - \ell_d| = 1 > 0$), it suffers more losses by self-imaging in the telescope (Fig. \ref{fig_degeneracylift}c at $\Delta z/4f = 1\%$), resulting in a donut with lower intensity than its neighbors (Fig. \ref{fig_defect}b). Finally, we note that multiple defects can be healed in the laser, provided that each defect is surrounded by a suitable neighbourhood as determined by the vortex coupling network (Fig. \ref{fig_L5}c).

%we consider that multiple defects can be healed in the laser when they randomly occur in the metasurface array with a probability up to $1/(N_c+1)$, where $N_c$ is the number of strongly coupled vortices. For $\ell_\mathrm{cr} = 1$, using the corresponding network shown in Fig. \ref{fig_L5}C ($N_c = 8$), this probability is estimated to be $1/9$---a value for which our simulations confirm defect healing.

\medskip
\textbf{\large{Discussion}}

We have demonstrated a metasurface laser capable of controlling topological interactions in optical vortex crystals based on arbitrary spin-orbit transformations. The underlying non-Hermitian mechanism provides stability and coherence to the system, as well as the ability to partition OAM within the cavity. The concepts presented in this work open up several new possibilities. The spatial coupling geometry of a laser array could be engineered by judiciously tailoring the topological charge of the beams. Anisotropic configurations could also be realized by employing arrays with fractional OAM. This could serve to individually control mutually interacting channels in a network of communicating nodes. We have focused here on the steady-state emission of optical vortex crystals from the laser, but in the future their charge could be modulated dynamically by adjusting the cavity length, either mechanically or optically. By exploiting the design freedom allowed by metasurfaces, many other topological interactions could be explored, such as non-reciprocal OAM transformations, where the imparted charge depends on which side of the metasurface is seen by the beam propagating through the cavity. The healing of defects more generally shows that the system can respond and adapt to aperiodic features of the phase mask. This opens up the possibility of iteratively manipulating information encoded as topological charges and could be used to develop photonic simulators \cite{Tradonsky2019} using OAM as a synthetic dimension \cite{Luo2015,PerezGarcia2018}. Finally, our scheme is scalable and allows for the generation of large optical vortex arrays that could exploit spatial-division multiplexing \cite{Richardson2013} for optical communications using multidimensional structured light \cite{piccardo2021roadmap} with OAM, frequency and polarization degrees of freedom.

\newpage

\textbf{\large{Methods}}

\small{\textbf{Metasurface arrays}---The metasurfaces consist of 600 nm-tall Si nanopillars arranged on an hexagonal close-packed lattice and lying on a 500 $\mu$m-thick fused SiO$_2$ substrate. The phase library of nanopillars was simulated for the Nd:YAG laser wavelength of 1064 nm using a finite-difference time-domain software (Lumerical) and a complex-value refractive index measured by ellipsometry. Each metasurface has a 200 $\mu$m diameter and is arranged in a 10$\times$10 square array with 300 $\mu$m period. For these array parameters the Talbot distance is $z_T = 169$ mm. The metasurfaces employ either geometric or propagation phase to impart a topological charge to the beam. The interstices between the metasurfaces are filled with a thin Au layer, which serves as an amplitude mask. The fabrication of the metasurface arrays includes the following main steps. A 600-nm-thick Si film, was deposited onto the fused SiO$_2$ substrates by plasma-enhanced chemical vapor deposition (STS LPX-PECVD system). Prior to e-beam exposure a 10-nm-thick Au film was deposited on top of the Poly(methyl methacrylate) layer by physical vapor deposition, thus ensuring for no charging effects. Electron beam direct-writing of the metasurface arrays was carried out using an ultra-high resolution Raith 150-Two e-beam lithography system. During exposure a beam energy of 20 keV was employed, setting the beam current to 140 pA. The Cr metasurface pattern obtained through lift-off was transferred to the Si film by means of a plasma enhanced reactive ion etching system (Sentech SI500). The etched depth and the sidewalls verticality were optimized by carefully tuning the process parameters (RF power 18 W, Pressure 0.32 Pa, ICP power 200 W, C$_4$F$_8$ 32 sccm, SF$_6$ 30 sccm, Ar 10 sccm, Temperature 0$^{\circ}$C). The amplitude mask was finally realized using standard UV lithography (Süss MicroTec MA6/BA6 mask aligner system) and chemical etching.

\textbf{Metasurface laser}---The active medium of the laser is an Nd:YAG module with 40 laser diode bars (Northrop Grumman-Cutting Edge Optronics). The Nd:YAG rod is low-doped ($<1\%$), 146 mm long and 7 mm in diameter. The module is water-cooled at 26$^{\circ}$C, and operated in quasi-continuous-wave mode with 250 $\mu$s current pulses at $1$ Hz repetition rate to avoid thermal lensing effects \cite{Pal2017}. Different intracavity optics schemes can be employed depending on the type of phase used in the metasurface design. We first align the resonator using a simple array of apertures (not generating a vortex crystal), and then we replace this by the array of metasurfaces. The intracavity lenses forming the telescope have an anti-reflection coating designed for the lasing wavelength, and a focal length of 150 mm. The output couplers have a reflectance above 97$\%$ (EKSMA Optics). The FF pinhole spatially filters the Fourier spectrum of the Gaussians making their intensity distribution more uniform and improving the quality of the vortices generated from the metasurface. At the same time the pinhole also blocks possible reflections from the metallic mask of the metasurface array. The FF pinhole could be removed if a non-reflective (non-metallic) amplitude mask is employed or when using $J$-plate metasurfaces in the vector vortex beam cavity configuration.

\textbf{Numerical simulations}---We carry out numerical simulations of the metasurface laser cavity based on the iterative Fox-Li method \cite{Fox1961}, including also the nonlinearity of the active medium via a gain matrix \cite{Tradonsky2017}. We start every simulation from a numerical noise seed to generate the initial complex field, which is then run through the cavity using the Fourier beam propagation method with an absorbing boundary layer \cite{Learn2016}. In particular, the propagation through the telescope section is not realized by means of Fourier transforms converting the beam from the near- to the far-field and viceversa, we rather implement phase matrices for the lenses and propagate the beam over the actual distances corresponding to our experimental setup. This allows calculating far-field patterns in scale with the experimental ones, as well as to account for the finite aperture of the intracavity optical elements. OAM transformations are accounted by implementing two different phase profiles on the opposite sides of the metasurface array, as those seen by the propagating beam.

\textbf{Topological charge analysis}---The setup for parallelized topological charge analysis consists of a $4f$ telescope with a phase-only SLM (HOLOEYE GAEA-2) lying at the Fourier plane in the middle of the two lenses. The SLM operates at first diffraction order in reflection mode, although in Fig. 2a it is shown in transmission mode for simplicity. As an initial charge analysis we decompose the beam using a pitchfork hologram, corresponding to the interference of a spiral phase with a diffraction grating (for the operation at first order). However, to correctly extract the relative weights among the different components of the topological spectrum, complex amplitude modulation is needed \cite{Pinnell2020}. This translates into an additional phase mask showing up as an annular element in the SLM holograms. To define the hologram used in the complex amplitude modulation we first fit the radius of the vortices imaged in the near-field of the metasurface array and extract the corresponding beam waist $\omega_0$ of the embedded Gaussian \cite{Sephton:16}. Then we simulate a single donut with the measured $\omega_0$ and the charge deduced from the pitchfork analysis, propagate it to the far-field and extract the waist of the corresponding far-field beam, which is used to program the SLM hologram. As the SLM only modulates horizontally-polarized light, the circularly-polarized light of the vortex crystal generated from $q$-plate arrays is converted to horizontally-polarized light using a combination of quarter- and half-waveplates (not shown in Fig. 2a). A pinhole is used to block unwanted diffraction orders before the camera (Spiricon LT665) acquisition. The modal decomposition of the array is carried out over a basis of Laguerre-Gaussian modes with topological charge ranging from $-2\ell$ to $2\ell$, where $\ell$ is the crystal charge, and $p=0$ radial mode. The optical axis of the setup was aligned by sending an array of Gaussian beams generated using a calibration mask of apertures instead of the metasurface array. The modal decomposition and the normalization of the topological charge spectra was carried out according to Ref. \cite{Pinnell2020}.
%For each individual beam, the topological charge spectrum was extracted by measuring the intensity at their optical axis for every detection mode and then normalized by their sum.
We employed background subtraction to remove noise from the acquired intensity images and set a noise threshold to exclude contributions of vortices that are too weak ($<2\%$) from the statistical average of the array. 

}

%\medskip
%\textbf{\normalsize{Data availability}}
%\small{The datasets needed to reproduce the figures of the main manuscript are provided in the online version. Additional data that support the findings of this study are available from the corresponding authors upon reasonable request.} 

\medskip
\textbf{\normalsize{Acknowledgements}}

This work has been financially supported by the European Research Council (ERC) under the European Union’s Horizon 2020 research and innovation programme ``METAmorphoses”, grant agreement no. 817794. This work has been supported by Fondazione Cariplo, grant no. 2019-3923.

%\medskip
%\textbf{\normalsize{Author contribution}}\newline
%\small{}

\medskip
\textbf{\normalsize{Competing interests}}\newline
\small{The authors declare no competing interests.}

\newpage

\textbf{An optical vortex crystal from the metasurface laser.} (\textbf{a}) Schematic of the laser cavity containing a metasurface array (MA) with charge $\ell_a$ that divides the resonator into two sections with different topological charges, $0$ and $\ell_a$. A $4f$ telescope with two lenses images the MA onto the left output coupler (OC). A pinhole forces the beams to carry zero topological charge in the telescope section. On the other side of the MA, the array of vortices are coupled by diffraction and form a self-image after a roundtrip distance of $z_T/2$ due to the Talbot effect. The polarization optics are specific to the metasurface design and exploit its engineered birefringence to convert the beam's spin-orbit state ($\alpha$ and $\beta$). (\textbf{b}) Optical microscope image of the MA ($\ell_a = 1$) acquired in transmission mode. The top inset shows a scanning electron microscope (SEM) image of engineered nanopillars. (\textbf{c}) Experimental near-field (NF) and far-field (FF) intensity distributions of the arrays emitted from opposite ends of the cavity, carrying a charge of 0 and 1. The Bragg diffraction patterns in the FF are a signature of the coherence of crystals with an out-of-phase relation. (\textbf{d}) Simulated Talbot carpet of an out-of-phase vortex crystal propagating along the $z$-direction. Five different transverse planes are shown, two of them (at $z_T/2$ and $z_T$) corresponding to self-images of the original array. (\textbf{e}) Convolution theorem applied to a vortex crystal that explains its FF pattern, which corresponds to a single pixelated donut. $\mathcal{F}$, Fourier transform. All scale bars are 300 $\mu$m, except for the SEM scale bar that is 500 nm.

\newpage

\textbf{Parallelized topological charge characterization of a vortex crystal.} (\textbf{a}) Each vortex in the array is projected to the same location in the far-field (FF) plane using a lens. In the FF a hologram encoded to modulate both phase and amplitude is displayed on a spatial light modulator (SLM) to carry out the charge analysis. A second lens re-images the entire array in the near-field (NF), where the new beam profiles carry information about the charge of the input vortices. The SLM operates in reflection with a first-order diffraction grating, but here it is shown in transmission and at zero order, for simplicity. (\textbf{b-e}) Topological charge spectra statistically averaged over the whole array for a metasurface array with a charge of one integrated in the laser. The different cases correspond to left-circular, right-circular, radial and azimuthal polarization outputs, respectively. (\textbf{f}) Topological charge decomposition of each vortex in the array for the laser with left-circular polarization (top). Also shown are the images of the modulated array obtained with different SLM holograms (bottom). The vortex crystal charge is identified when the image corresponds to an array of Gaussian beams.

\newpage

\newpage

\textbf{Partitioning orbital angular momentum inside the cavity.} (\textbf{a}) The partitioning of OAM in the cavity originates from the spatial filtering of the mask following a roundtrip in either the telescope or Talbot section of the cavity, which constitute OAM-dependent losses. The first type of loss depends on the cavity adjustment $\Delta z$ with respect to the $4f$ value. Also shown are experimental images of representative beams extracted from two crystals emitted from the same side of the metasurface laser (array charge $\ell_a = 1$) for different $\Delta z/4f$, accompanied by the charge spectra of the respective crystals. (\textbf{b}) Regions of existence of all the $\ell_a+1$ topological solutions as a function of the metasurface array charge and $\Delta z/4f$. The solutions are written as $(m~|~n)$, where $m$ and $n$ are the charges of the crystals on the telescope and Talbot section of the cavity, respectively. (\textbf{c}, \textbf{d}) Calculated transmission efficiency of a $10\times10$ array of vortex beams filtered by the mask after a roundtrip through: (\textbf{c}) the telescope section; (\textbf{d}) the Talbot section. (\textbf{e}) The combination of the two calculated OAM-dependent losses determines which topological solution is favored by the laser as a function of $\Delta z$, corresponding to the curve with the largest transmission. The case of $\ell_a = 1$ is shown.

\newpage

\begin{thebibliography}{52}%
\makeatletter
\providecommand \@ifxundefined [1]{%
 \@ifx{#1\undefined}
}%
\providecommand \@ifnum [1]{%
 \ifnum #1\expandafter \@firstoftwo
 \else \expandafter \@secondoftwo
 \fi
}%
\providecommand \@ifx [1]{%
 \ifx #1\expandafter \@firstoftwo
 \else \expandafter \@secondoftwo
 \fi
}%
\providecommand \natexlab [1]{#1}%
\providecommand \enquote  [1]{``#1''}%
\providecommand \bibnamefont  [1]{#1}%
\providecommand \bibfnamefont [1]{#1}%
\providecommand \citenamefont [1]{#1}%
\providecommand \href@noop [0]{\@secondoftwo}%
\providecommand \href [0]{\begingroup \@sanitize@url \@href}%
\providecommand \@href[1]{\@@startlink{#1}\@@href}%
\providecommand \@@href[1]{\endgroup#1\@@endlink}%
\providecommand \@sanitize@url [0]{\catcode `\\12\catcode `\$12\catcode
  `\&12\catcode `\#12\catcode `\^12\catcode `\_12\catcode `\%12\relax}%
\providecommand \@@startlink[1]{}%
\providecommand \@@endlink[0]{}%
\providecommand \url  [0]{\begingroup\@sanitize@url \@url }%
\providecommand \@url [1]{\endgroup\@href {#1}{\urlprefix }}%
\providecommand \urlprefix  [0]{URL }%
\providecommand \Eprint [0]{\href }%
\providecommand \doibase [0]{http://dx.doi.org/}%
\providecommand \selectlanguage [0]{\@gobble}%
\providecommand \bibinfo  [0]{\@secondoftwo}%
\providecommand \bibfield  [0]{\@secondoftwo}%
\providecommand \translation [1]{[#1]}%
\providecommand \BibitemOpen [0]{}%
\providecommand \bibitemStop [0]{}%
\providecommand \bibitemNoStop [0]{.\EOS\space}%
\providecommand \EOS [0]{\spacefactor3000\relax}%
\providecommand \BibitemShut  [1]{\csname bibitem#1\endcsname}%
\let\auto@bib@innerbib\@empty
%</preamble>
\bibitem [{\citenamefont {Jin}\ and\ \citenamefont {Dubin}(2000)}]{Jin2000}%
  \BibitemOpen
  \bibfield  {author} {\bibinfo {author} {\bibfnamefont {D.~Z.}\ \bibnamefont
  {Jin}}\ and\ \bibinfo {author} {\bibfnamefont {D.~H.~E.}\ \bibnamefont
  {Dubin}},\ }\href {\doibase 10.1103/PhysRevLett.84.1443} {\bibfield
  {journal} {\bibinfo  {journal} {PRL}\ }\textbf {\bibinfo {volume} {84}},\
  \bibinfo {pages} {1443} (\bibinfo {year} {2000})}\BibitemShut {NoStop}%
\bibitem [{\citenamefont {Adriani}\ \emph {et~al.}(2018)\citenamefont
  {Adriani}, \citenamefont {Mura}, \citenamefont {Orton}, \citenamefont
  {Hansen}, \citenamefont {Altieri}, \citenamefont {Moriconi}, \citenamefont
  {Rogers}, \citenamefont {Eichstädt}, \citenamefont {Momary}, \citenamefont
  {Ingersoll}, \citenamefont {Filacchione}, \citenamefont {Sindoni},
  \citenamefont {Tabataba-Vakili}, \citenamefont {Dinelli}, \citenamefont
  {Fabiano}, \citenamefont {Bolton}, \citenamefont {Connerney}, \citenamefont
  {Atreya}, \citenamefont {Lunine}, \citenamefont {Tosi}, \citenamefont
  {Migliorini}, \citenamefont {Grassi}, \citenamefont {Piccioni}, \citenamefont
  {Noschese}, \citenamefont {Cicchetti}, \citenamefont {Plainaki},
  \citenamefont {Olivieri}, \citenamefont {O’Neill}, \citenamefont {Turrini},
  \citenamefont {Stefani}, \citenamefont {Sordini},\ and\ \citenamefont
  {Amoroso}}]{Adriani2018}%
  \BibitemOpen
  \bibfield  {author} {\bibinfo {author} {\bibfnamefont {A.}~\bibnamefont
  {Adriani}}, \bibinfo {author} {\bibfnamefont {A.}~\bibnamefont {Mura}},
  \bibinfo {author} {\bibfnamefont {G.}~\bibnamefont {Orton}}, \bibinfo
  {author} {\bibfnamefont {C.}~\bibnamefont {Hansen}}, \bibinfo {author}
  {\bibfnamefont {F.}~\bibnamefont {Altieri}}, \bibinfo {author} {\bibfnamefont
  {M.~L.}\ \bibnamefont {Moriconi}}, \bibinfo {author} {\bibfnamefont
  {J.}~\bibnamefont {Rogers}}, \bibinfo {author} {\bibfnamefont
  {G.}~\bibnamefont {Eichstädt}}, \bibinfo {author} {\bibfnamefont
  {T.}~\bibnamefont {Momary}}, \bibinfo {author} {\bibfnamefont {A.~P.}\
  \bibnamefont {Ingersoll}}, \bibinfo {author} {\bibfnamefont {G.}~\bibnamefont
  {Filacchione}}, \bibinfo {author} {\bibfnamefont {G.}~\bibnamefont
  {Sindoni}}, \bibinfo {author} {\bibfnamefont {F.}~\bibnamefont
  {Tabataba-Vakili}}, \bibinfo {author} {\bibfnamefont {B.~M.}\ \bibnamefont
  {Dinelli}}, \bibinfo {author} {\bibfnamefont {F.}~\bibnamefont {Fabiano}},
  \bibinfo {author} {\bibfnamefont {S.~J.}\ \bibnamefont {Bolton}}, \bibinfo
  {author} {\bibfnamefont {J.~E.~P.}\ \bibnamefont {Connerney}}, \bibinfo
  {author} {\bibfnamefont {S.~K.}\ \bibnamefont {Atreya}}, \bibinfo {author}
  {\bibfnamefont {J.~I.}\ \bibnamefont {Lunine}}, \bibinfo {author}
  {\bibfnamefont {F.}~\bibnamefont {Tosi}}, \bibinfo {author} {\bibfnamefont
  {A.}~\bibnamefont {Migliorini}}, \bibinfo {author} {\bibfnamefont
  {D.}~\bibnamefont {Grassi}}, \bibinfo {author} {\bibfnamefont
  {G.}~\bibnamefont {Piccioni}}, \bibinfo {author} {\bibfnamefont
  {R.}~\bibnamefont {Noschese}}, \bibinfo {author} {\bibfnamefont
  {A.}~\bibnamefont {Cicchetti}}, \bibinfo {author} {\bibfnamefont
  {C.}~\bibnamefont {Plainaki}}, \bibinfo {author} {\bibfnamefont
  {A.}~\bibnamefont {Olivieri}}, \bibinfo {author} {\bibfnamefont {M.~E.}\
  \bibnamefont {O’Neill}}, \bibinfo {author} {\bibfnamefont {D.}~\bibnamefont
  {Turrini}}, \bibinfo {author} {\bibfnamefont {S.}~\bibnamefont {Stefani}},
  \bibinfo {author} {\bibfnamefont {R.}~\bibnamefont {Sordini}}, \ and\
  \bibinfo {author} {\bibfnamefont {M.}~\bibnamefont {Amoroso}},\ }\href
  {\doibase 10.1038/nature25491} {\bibfield  {journal} {\bibinfo  {journal}
  {Nature}\ }\textbf {\bibinfo {volume} {555}},\ \bibinfo {pages} {216}
  (\bibinfo {year} {2018})}\BibitemShut {NoStop}%
\bibitem [{\citenamefont {Tai}\ and\ \citenamefont {Smalyukh}(2019)}]{Tai2019}%
  \BibitemOpen
  \bibfield  {author} {\bibinfo {author} {\bibfnamefont {J.-S.~B.}\
  \bibnamefont {Tai}}\ and\ \bibinfo {author} {\bibfnamefont {I.~I.}\
  \bibnamefont {Smalyukh}},\ }\href {\doibase 10.1126/science.aay1638}
  {\bibfield  {journal} {\bibinfo  {journal} {Science}\ }\textbf {\bibinfo
  {volume} {365}},\ \bibinfo {pages} {1449} (\bibinfo {year}
  {2019})}\BibitemShut {NoStop}%
\bibitem [{\citenamefont {Riedel}\ \emph {et~al.}(2005)\citenamefont {Riedel},
  \citenamefont {Kruse},\ and\ \citenamefont {Howard}}]{Riedel2005}%
  \BibitemOpen
  \bibfield  {author} {\bibinfo {author} {\bibfnamefont {I.~H.}\ \bibnamefont
  {Riedel}}, \bibinfo {author} {\bibfnamefont {K.}~\bibnamefont {Kruse}}, \
  and\ \bibinfo {author} {\bibfnamefont {J.}~\bibnamefont {Howard}},\ }\href
  {\doibase 10.1126/science.1110329} {\bibfield  {journal} {\bibinfo  {journal}
  {Science}\ }\textbf {\bibinfo {volume} {309}},\ \bibinfo {pages} {300}
  (\bibinfo {year} {2005})}\BibitemShut {NoStop}%
\bibitem [{\citenamefont {Brambilla}\ \emph {et~al.}(1991)\citenamefont
  {Brambilla}, \citenamefont {Battipede}, \citenamefont {Lugiato},
  \citenamefont {Penna}, \citenamefont {Prati}, \citenamefont {Tamm},\ and\
  \citenamefont {Weiss}}]{Brambilla1991}%
  \BibitemOpen
  \bibfield  {author} {\bibinfo {author} {\bibfnamefont {M.}~\bibnamefont
  {Brambilla}}, \bibinfo {author} {\bibfnamefont {F.}~\bibnamefont
  {Battipede}}, \bibinfo {author} {\bibfnamefont {L.~A.}\ \bibnamefont
  {Lugiato}}, \bibinfo {author} {\bibfnamefont {V.}~\bibnamefont {Penna}},
  \bibinfo {author} {\bibfnamefont {F.}~\bibnamefont {Prati}}, \bibinfo
  {author} {\bibfnamefont {C.}~\bibnamefont {Tamm}}, \ and\ \bibinfo {author}
  {\bibfnamefont {C.~O.}\ \bibnamefont {Weiss}},\ }\href {\doibase
  10.1103/PhysRevA.43.5090} {\bibfield  {journal} {\bibinfo  {journal} {Phys.
  Rev. A}\ }\textbf {\bibinfo {volume} {43}},\ \bibinfo {pages} {5090}
  (\bibinfo {year} {1991})}\BibitemShut {NoStop}%
\bibitem [{\citenamefont {Scheuer}\ and\ \citenamefont
  {Orenstein}(1999)}]{Scheuer1999}%
  \BibitemOpen
  \bibfield  {author} {\bibinfo {author} {\bibfnamefont {J.}~\bibnamefont
  {Scheuer}}\ and\ \bibinfo {author} {\bibfnamefont {M.}~\bibnamefont
  {Orenstein}},\ }\href {\doibase 10.1126/science.285.5425.230} {\bibfield
  {journal} {\bibinfo  {journal} {Science}\ }\textbf {\bibinfo {volume}
  {285}},\ \bibinfo {pages} {230} (\bibinfo {year} {1999})}\BibitemShut
  {NoStop}%
\bibitem [{\citenamefont {Coullet}\ \emph {et~al.}(1989)\citenamefont
  {Coullet}, \citenamefont {Gil},\ and\ \citenamefont {Rocca}}]{Coullet1989}%
  \BibitemOpen
  \bibfield  {author} {\bibinfo {author} {\bibfnamefont {P.}~\bibnamefont
  {Coullet}}, \bibinfo {author} {\bibfnamefont {L.}~\bibnamefont {Gil}}, \ and\
  \bibinfo {author} {\bibfnamefont {F.}~\bibnamefont {Rocca}},\ }\href
  {http://www.sciencedirect.com/science/article/pii/0030401889901806}
  {\bibfield  {journal} {\bibinfo  {journal} {Optics Communications}\ }\textbf
  {\bibinfo {volume} {73}},\ \bibinfo {pages} {403} (\bibinfo {year}
  {1989})}\BibitemShut {NoStop}%
\bibitem [{\citenamefont {Allen}\ \emph {et~al.}(1992)\citenamefont {Allen},
  \citenamefont {Beijersbergen}, \citenamefont {Spreeuw},\ and\ \citenamefont
  {Woerdman}}]{Allen1992}%
  \BibitemOpen
  \bibfield  {author} {\bibinfo {author} {\bibfnamefont {L.}~\bibnamefont
  {Allen}}, \bibinfo {author} {\bibfnamefont {M.~W.}\ \bibnamefont
  {Beijersbergen}}, \bibinfo {author} {\bibfnamefont {R.~J.~C.}\ \bibnamefont
  {Spreeuw}}, \ and\ \bibinfo {author} {\bibfnamefont {J.~P.}\ \bibnamefont
  {Woerdman}},\ }\href@noop {} {\bibfield  {journal} {\bibinfo  {journal}
  {PRA}\ }\textbf {\bibinfo {volume} {45}},\ \bibinfo {pages} {8185} (\bibinfo
  {year} {1992})}\BibitemShut {NoStop}%
\bibitem [{\citenamefont {Staliunas}\ and\ \citenamefont
  {Sanchez-Morcillo}(2003)}]{staliunas2003transverse}%
  \BibitemOpen
  \bibfield  {author} {\bibinfo {author} {\bibfnamefont {K.}~\bibnamefont
  {Staliunas}}\ and\ \bibinfo {author} {\bibfnamefont {V.~J.}\ \bibnamefont
  {Sanchez-Morcillo}},\ }\href@noop {} {\bibfield  {journal} {\bibinfo
  {journal} {Springer}\ } (\bibinfo {year} {2003})}\BibitemShut {NoStop}%
\bibitem [{\citenamefont {Wang}\ \emph {et~al.}(2018)\citenamefont {Wang},
  \citenamefont {Nie}, \citenamefont {Liang}, \citenamefont {Wang},
  \citenamefont {Li},\ and\ \citenamefont {Jia}}]{Wang2018}%
  \BibitemOpen
  \bibfield  {author} {\bibinfo {author} {\bibfnamefont {X.}~\bibnamefont
  {Wang}}, \bibinfo {author} {\bibfnamefont {Z.}~\bibnamefont {Nie}}, \bibinfo
  {author} {\bibfnamefont {Y.}~\bibnamefont {Liang}}, \bibinfo {author}
  {\bibfnamefont {J.}~\bibnamefont {Wang}}, \bibinfo {author} {\bibfnamefont
  {T.}~\bibnamefont {Li}}, \ and\ \bibinfo {author} {\bibfnamefont
  {B.}~\bibnamefont {Jia}},\ }\href@noop {} {\bibfield  {journal} {\bibinfo
  {journal} {Nanophotonics}\ }\textbf {\bibinfo {volume} {7}},\ \bibinfo
  {pages} {1533} (\bibinfo {year} {2018})}\BibitemShut {NoStop}%
\bibitem [{\citenamefont {Shen}\ \emph {et~al.}(2019)\citenamefont {Shen},
  \citenamefont {Wang}, \citenamefont {Xie}, \citenamefont {Min}, \citenamefont
  {Fu}, \citenamefont {Liu}, \citenamefont {Gong},\ and\ \citenamefont
  {Yuan}}]{Shen2019}%
  \BibitemOpen
  \bibfield  {author} {\bibinfo {author} {\bibfnamefont {Y.}~\bibnamefont
  {Shen}}, \bibinfo {author} {\bibfnamefont {X.}~\bibnamefont {Wang}}, \bibinfo
  {author} {\bibfnamefont {Z.}~\bibnamefont {Xie}}, \bibinfo {author}
  {\bibfnamefont {C.}~\bibnamefont {Min}}, \bibinfo {author} {\bibfnamefont
  {X.}~\bibnamefont {Fu}}, \bibinfo {author} {\bibfnamefont {Q.}~\bibnamefont
  {Liu}}, \bibinfo {author} {\bibfnamefont {M.}~\bibnamefont {Gong}}, \ and\
  \bibinfo {author} {\bibfnamefont {X.}~\bibnamefont {Yuan}},\ }\href {\doibase
  10.1038/s41377-019-0194-2} {\bibfield  {journal} {\bibinfo  {journal} {Light:
  Science \& Applications}\ }\textbf {\bibinfo {volume} {8}},\ \bibinfo {pages}
  {90} (\bibinfo {year} {2019})}\BibitemShut {NoStop}%
\bibitem [{\citenamefont {Vicidomini}\ \emph {et~al.}(2018)\citenamefont
  {Vicidomini}, \citenamefont {Bianchini},\ and\ \citenamefont
  {Diaspro}}]{Vicidomini2018}%
  \BibitemOpen
  \bibfield  {author} {\bibinfo {author} {\bibfnamefont {G.}~\bibnamefont
  {Vicidomini}}, \bibinfo {author} {\bibfnamefont {P.}~\bibnamefont
  {Bianchini}}, \ and\ \bibinfo {author} {\bibfnamefont {A.}~\bibnamefont
  {Diaspro}},\ }\href {\doibase 10.1038/nmeth.4593} {\bibfield  {journal}
  {\bibinfo  {journal} {Nature Methods}\ }\textbf {\bibinfo {volume} {15}},\
  \bibinfo {pages} {173} (\bibinfo {year} {2018})}\BibitemShut {NoStop}%
\bibitem [{\citenamefont {Ren}\ \emph {et~al.}(2016)\citenamefont {Ren},
  \citenamefont {Li}, \citenamefont {Zhang},\ and\ \citenamefont
  {Gu}}]{Ren2016}%
  \BibitemOpen
  \bibfield  {author} {\bibinfo {author} {\bibfnamefont {H.}~\bibnamefont
  {Ren}}, \bibinfo {author} {\bibfnamefont {X.}~\bibnamefont {Li}}, \bibinfo
  {author} {\bibfnamefont {Q.}~\bibnamefont {Zhang}}, \ and\ \bibinfo {author}
  {\bibfnamefont {M.}~\bibnamefont {Gu}},\ }\href {\doibase
  10.1126/science.aaf1112} {\bibfield  {journal} {\bibinfo  {journal}
  {Science}\ }\textbf {\bibinfo {volume} {352}},\ \bibinfo {pages} {805}
  (\bibinfo {year} {2016})}\BibitemShut {NoStop}%
\bibitem [{\citenamefont {Padgett}\ and\ \citenamefont
  {Bowman}(2011)}]{Padgett2011}%
  \BibitemOpen
  \bibfield  {author} {\bibinfo {author} {\bibfnamefont {M.}~\bibnamefont
  {Padgett}}\ and\ \bibinfo {author} {\bibfnamefont {R.}~\bibnamefont
  {Bowman}},\ }\href {\doibase 10.1038/nphoton.2011.81} {\bibfield  {journal}
  {\bibinfo  {journal} {Nature Photonics}\ }\textbf {\bibinfo {volume} {5}},\
  \bibinfo {pages} {343} (\bibinfo {year} {2011})}\BibitemShut {NoStop}%
\bibitem [{\citenamefont {Woerdemann}\ \emph {et~al.}(2013)\citenamefont
  {Woerdemann}, \citenamefont {Alpmann}, \citenamefont {Esseling},\ and\
  \citenamefont {Denz}}]{Woerdemann2013}%
  \BibitemOpen
  \bibfield  {author} {\bibinfo {author} {\bibfnamefont {M.}~\bibnamefont
  {Woerdemann}}, \bibinfo {author} {\bibfnamefont {C.}~\bibnamefont {Alpmann}},
  \bibinfo {author} {\bibfnamefont {M.}~\bibnamefont {Esseling}}, \ and\
  \bibinfo {author} {\bibfnamefont {C.}~\bibnamefont {Denz}},\ }\href {\doibase
  https://doi.org/10.1002/lpor.201200058} {\bibfield  {journal} {\bibinfo
  {journal} {Laser \& Photonics Reviews}\ }\textbf {\bibinfo {volume} {7}},\
  \bibinfo {pages} {839} (\bibinfo {year} {2013})}\BibitemShut {NoStop}%
\bibitem [{\citenamefont {Lei}\ \emph {et~al.}(2015)\citenamefont {Lei},
  \citenamefont {Zhang}, \citenamefont {Li}, \citenamefont {Jia}, \citenamefont
  {Liu}, \citenamefont {Xu}, \citenamefont {Li}, \citenamefont {Min},
  \citenamefont {Lin}, \citenamefont {Yu}, \citenamefont {Niu},\ and\
  \citenamefont {Yuan}}]{Lei2015}%
  \BibitemOpen
  \bibfield  {author} {\bibinfo {author} {\bibfnamefont {T.}~\bibnamefont
  {Lei}}, \bibinfo {author} {\bibfnamefont {M.}~\bibnamefont {Zhang}}, \bibinfo
  {author} {\bibfnamefont {Y.}~\bibnamefont {Li}}, \bibinfo {author}
  {\bibfnamefont {P.}~\bibnamefont {Jia}}, \bibinfo {author} {\bibfnamefont
  {G.~N.}\ \bibnamefont {Liu}}, \bibinfo {author} {\bibfnamefont
  {X.}~\bibnamefont {Xu}}, \bibinfo {author} {\bibfnamefont {Z.}~\bibnamefont
  {Li}}, \bibinfo {author} {\bibfnamefont {C.}~\bibnamefont {Min}}, \bibinfo
  {author} {\bibfnamefont {J.}~\bibnamefont {Lin}}, \bibinfo {author}
  {\bibfnamefont {C.}~\bibnamefont {Yu}}, \bibinfo {author} {\bibfnamefont
  {H.}~\bibnamefont {Niu}}, \ and\ \bibinfo {author} {\bibfnamefont
  {X.}~\bibnamefont {Yuan}},\ }\href {\doibase 10.1038/lsa.2015.30} {\bibfield
  {journal} {\bibinfo  {journal} {Light: Science \& Applications}\ }\textbf
  {\bibinfo {volume} {4}},\ \bibinfo {pages} {257} (\bibinfo {year}
  {2015})}\BibitemShut {NoStop}%
\bibitem [{\citenamefont {Kildishev}\ \emph {et~al.}(2013)\citenamefont
  {Kildishev}, \citenamefont {Boltasseva},\ and\ \citenamefont
  {Shalaev}}]{Kildishev2013}%
  \BibitemOpen
  \bibfield  {author} {\bibinfo {author} {\bibfnamefont {A.~V.}\ \bibnamefont
  {Kildishev}}, \bibinfo {author} {\bibfnamefont {A.}~\bibnamefont
  {Boltasseva}}, \ and\ \bibinfo {author} {\bibfnamefont {V.~M.}\ \bibnamefont
  {Shalaev}},\ }\href {\doibase 10.1126/science.1232009} {\bibfield  {journal}
  {\bibinfo  {journal} {Science}\ }\textbf {\bibinfo {volume} {339}},\ \bibinfo
  {pages} {1232009} (\bibinfo {year} {2013})}\BibitemShut {NoStop}%
\bibitem [{\citenamefont {Yu}\ and\ \citenamefont {Capasso}(2014)}]{Yu2014}%
  \BibitemOpen
  \bibfield  {author} {\bibinfo {author} {\bibfnamefont {N.}~\bibnamefont
  {Yu}}\ and\ \bibinfo {author} {\bibfnamefont {F.}~\bibnamefont {Capasso}},\
  }\href {\doibase 10.1038/nmat3839} {\bibfield  {journal} {\bibinfo  {journal}
  {Nature Materials}\ }\textbf {\bibinfo {volume} {13}},\ \bibinfo {pages}
  {139} (\bibinfo {year} {2014})}\BibitemShut {NoStop}%
\bibitem [{\citenamefont {Maguid}\ \emph {et~al.}(2018)\citenamefont {Maguid},
  \citenamefont {Chriki}, \citenamefont {Yannai}, \citenamefont {Kleiner},
  \citenamefont {Hasman}, \citenamefont {Friesem},\ and\ \citenamefont
  {Davidson}}]{Maguid2018}%
  \BibitemOpen
  \bibfield  {author} {\bibinfo {author} {\bibfnamefont {E.}~\bibnamefont
  {Maguid}}, \bibinfo {author} {\bibfnamefont {R.}~\bibnamefont {Chriki}},
  \bibinfo {author} {\bibfnamefont {M.}~\bibnamefont {Yannai}}, \bibinfo
  {author} {\bibfnamefont {V.}~\bibnamefont {Kleiner}}, \bibinfo {author}
  {\bibfnamefont {E.}~\bibnamefont {Hasman}}, \bibinfo {author} {\bibfnamefont
  {A.~A.}\ \bibnamefont {Friesem}}, \ and\ \bibinfo {author} {\bibfnamefont
  {N.}~\bibnamefont {Davidson}},\ }\href {\doibase
  10.1021/acsphotonics.7b01525} {\bibfield  {journal} {\bibinfo  {journal} {ACS
  Photonics}\ }\textbf {\bibinfo {volume} {5}},\ \bibinfo {pages} {1817}
  (\bibinfo {year} {2018})}\BibitemShut {NoStop}%
\bibitem [{\citenamefont {Jin}\ \emph {et~al.}(2017)\citenamefont {Jin},
  \citenamefont {Pu}, \citenamefont {Wang}, \citenamefont {Li}, \citenamefont
  {Ma}, \citenamefont {Luo}, \citenamefont {Zhao}, \citenamefont {Gao},\ and\
  \citenamefont {Luo}}]{Jin2017}%
  \BibitemOpen
  \bibfield  {author} {\bibinfo {author} {\bibfnamefont {J.}~\bibnamefont
  {Jin}}, \bibinfo {author} {\bibfnamefont {M.}~\bibnamefont {Pu}}, \bibinfo
  {author} {\bibfnamefont {Y.}~\bibnamefont {Wang}}, \bibinfo {author}
  {\bibfnamefont {X.}~\bibnamefont {Li}}, \bibinfo {author} {\bibfnamefont
  {X.}~\bibnamefont {Ma}}, \bibinfo {author} {\bibfnamefont {J.}~\bibnamefont
  {Luo}}, \bibinfo {author} {\bibfnamefont {Z.}~\bibnamefont {Zhao}}, \bibinfo
  {author} {\bibfnamefont {P.}~\bibnamefont {Gao}}, \ and\ \bibinfo {author}
  {\bibfnamefont {X.}~\bibnamefont {Luo}},\ }\href {\doibase
  https://doi.org/10.1002/admt.201600201} {\bibfield  {journal} {\bibinfo
  {journal} {Advanced Materials Technologies}\ }\textbf {\bibinfo {volume}
  {2}},\ \bibinfo {pages} {1600201} (\bibinfo {year} {2017})}\BibitemShut
  {NoStop}%
\bibitem [{\citenamefont {Piccardo}\ and\ \citenamefont
  {Ambrosio}(2020)}]{Piccardo2020APL}%
  \BibitemOpen
  \bibfield  {author} {\bibinfo {author} {\bibfnamefont {M.}~\bibnamefont
  {Piccardo}}\ and\ \bibinfo {author} {\bibfnamefont {A.}~\bibnamefont
  {Ambrosio}},\ }\href {\doibase 10.1063/5.0023338} {\bibfield  {journal}
  {\bibinfo  {journal} {Applied Physics Letters}\ }\textbf {\bibinfo {volume}
  {117}},\ \bibinfo {pages} {140501} (\bibinfo {year} {2020})}\BibitemShut
  {NoStop}%
\bibitem [{\citenamefont {Qiao}\ \emph {et~al.}(2021)\citenamefont {Qiao},
  \citenamefont {Midya}, \citenamefont {Gao}, \citenamefont {Zhang},
  \citenamefont {Zhao}, \citenamefont {Wu}, \citenamefont {Yim}, \citenamefont
  {Agarwal}, \citenamefont {Litchinitser},\ and\ \citenamefont
  {Feng}}]{Qiao2021}%
  \BibitemOpen
  \bibfield  {author} {\bibinfo {author} {\bibfnamefont {X.}~\bibnamefont
  {Qiao}}, \bibinfo {author} {\bibfnamefont {B.}~\bibnamefont {Midya}},
  \bibinfo {author} {\bibfnamefont {Z.}~\bibnamefont {Gao}}, \bibinfo {author}
  {\bibfnamefont {Z.}~\bibnamefont {Zhang}}, \bibinfo {author} {\bibfnamefont
  {H.}~\bibnamefont {Zhao}}, \bibinfo {author} {\bibfnamefont {T.}~\bibnamefont
  {Wu}}, \bibinfo {author} {\bibfnamefont {J.}~\bibnamefont {Yim}}, \bibinfo
  {author} {\bibfnamefont {R.}~\bibnamefont {Agarwal}}, \bibinfo {author}
  {\bibfnamefont {N.~M.}\ \bibnamefont {Litchinitser}}, \ and\ \bibinfo
  {author} {\bibfnamefont {L.}~\bibnamefont {Feng}},\ }\href {\doibase
  10.1126/science.abg3904} {\bibfield  {journal} {\bibinfo  {journal}
  {Science}\ }\textbf {\bibinfo {volume} {372}},\ \bibinfo {pages} {403}
  (\bibinfo {year} {2021})}\BibitemShut {NoStop}%
\bibitem [{\citenamefont {Forbes}(2019)}]{Forbes2019}%
  \BibitemOpen
  \bibfield  {author} {\bibinfo {author} {\bibfnamefont {A.}~\bibnamefont
  {Forbes}},\ }\href {\doibase https://doi.org/10.1002/lpor.201900140}
  {\bibfield  {journal} {\bibinfo  {journal} {Laser \& Photonics Reviews}\
  }\textbf {\bibinfo {volume} {13}},\ \bibinfo {pages} {1900140} (\bibinfo
  {year} {2019})}\BibitemShut {NoStop}%
\bibitem [{\citenamefont {Nixon}\ \emph
  {et~al.}(2013{\natexlab{a}})\citenamefont {Nixon}, \citenamefont {Ronen},
  \citenamefont {Friesem},\ and\ \citenamefont {Davidson}}]{Nixon2013}%
  \BibitemOpen
  \bibfield  {author} {\bibinfo {author} {\bibfnamefont {M.}~\bibnamefont
  {Nixon}}, \bibinfo {author} {\bibfnamefont {E.}~\bibnamefont {Ronen}},
  \bibinfo {author} {\bibfnamefont {A.~A.}\ \bibnamefont {Friesem}}, \ and\
  \bibinfo {author} {\bibfnamefont {N.}~\bibnamefont {Davidson}},\ }\href
  {\doibase 10.1103/PhysRevLett.110.184102} {\bibfield  {journal} {\bibinfo
  {journal} {Phys. Rev. Lett.}\ }\textbf {\bibinfo {volume} {110}},\ \bibinfo
  {pages} {184102} (\bibinfo {year} {2013}{\natexlab{a}})}\BibitemShut
  {NoStop}%
\bibitem [{\citenamefont {Pal}\ \emph {et~al.}(2017)\citenamefont {Pal},
  \citenamefont {Tradonsky}, \citenamefont {Chriki}, \citenamefont {Friesem},\
  and\ \citenamefont {Davidson}}]{Pal2017}%
  \BibitemOpen
  \bibfield  {author} {\bibinfo {author} {\bibfnamefont {V.}~\bibnamefont
  {Pal}}, \bibinfo {author} {\bibfnamefont {C.}~\bibnamefont {Tradonsky}},
  \bibinfo {author} {\bibfnamefont {R.}~\bibnamefont {Chriki}}, \bibinfo
  {author} {\bibfnamefont {A.~A.}\ \bibnamefont {Friesem}}, \ and\ \bibinfo
  {author} {\bibfnamefont {N.}~\bibnamefont {Davidson}},\ }\href {\doibase
  10.1103/PhysRevLett.119.013902} {\bibfield  {journal} {\bibinfo  {journal}
  {Phys. Rev. Lett.}\ }\textbf {\bibinfo {volume} {119}},\ \bibinfo {pages}
  {013902} (\bibinfo {year} {2017})}\BibitemShut {NoStop}%
\bibitem [{\citenamefont {Naidoo}\ \emph {et~al.}(2016)\citenamefont {Naidoo},
  \citenamefont {Roux}, \citenamefont {Dudley}, \citenamefont {Litvin},
  \citenamefont {Piccirillo}, \citenamefont {Marrucci},\ and\ \citenamefont
  {Forbes}}]{Naidoo2016}%
  \BibitemOpen
  \bibfield  {author} {\bibinfo {author} {\bibfnamefont {D.}~\bibnamefont
  {Naidoo}}, \bibinfo {author} {\bibfnamefont {F.~S.}\ \bibnamefont {Roux}},
  \bibinfo {author} {\bibfnamefont {A.}~\bibnamefont {Dudley}}, \bibinfo
  {author} {\bibfnamefont {I.}~\bibnamefont {Litvin}}, \bibinfo {author}
  {\bibfnamefont {B.}~\bibnamefont {Piccirillo}}, \bibinfo {author}
  {\bibfnamefont {L.}~\bibnamefont {Marrucci}}, \ and\ \bibinfo {author}
  {\bibfnamefont {A.}~\bibnamefont {Forbes}},\ }\href {\doibase
  10.1038/nphoton.2016.37} {\bibfield  {journal} {\bibinfo  {journal} {Nature
  Photonics}\ }\textbf {\bibinfo {volume} {10}},\ \bibinfo {pages} {327}
  (\bibinfo {year} {2016})}\BibitemShut {NoStop}%
\bibitem [{\citenamefont {Sroor}\ \emph {et~al.}(2020)\citenamefont {Sroor},
  \citenamefont {Huang}, \citenamefont {Sephton}, \citenamefont {Naidoo},
  \citenamefont {Vallés}, \citenamefont {Ginis}, \citenamefont {Qiu},
  \citenamefont {Ambrosio}, \citenamefont {Capasso},\ and\ \citenamefont
  {Forbes}}]{Sroor2020}%
  \BibitemOpen
  \bibfield  {author} {\bibinfo {author} {\bibfnamefont {H.}~\bibnamefont
  {Sroor}}, \bibinfo {author} {\bibfnamefont {Y.-W.}\ \bibnamefont {Huang}},
  \bibinfo {author} {\bibfnamefont {B.}~\bibnamefont {Sephton}}, \bibinfo
  {author} {\bibfnamefont {D.}~\bibnamefont {Naidoo}}, \bibinfo {author}
  {\bibfnamefont {A.}~\bibnamefont {Vallés}}, \bibinfo {author} {\bibfnamefont
  {V.}~\bibnamefont {Ginis}}, \bibinfo {author} {\bibfnamefont {C.-W.}\
  \bibnamefont {Qiu}}, \bibinfo {author} {\bibfnamefont {A.}~\bibnamefont
  {Ambrosio}}, \bibinfo {author} {\bibfnamefont {F.}~\bibnamefont {Capasso}}, \
  and\ \bibinfo {author} {\bibfnamefont {A.}~\bibnamefont {Forbes}},\
  }\href@noop {} {\bibfield  {journal} {\bibinfo  {journal} {Nature Photonics}\
  }\textbf {\bibinfo {volume} {14}},\ \bibinfo {pages} {498} (\bibinfo {year}
  {2020})}\BibitemShut {NoStop}%
\bibitem [{\citenamefont {Ito}\ \emph {et~al.}(2010)\citenamefont {Ito},
  \citenamefont {Kozawa},\ and\ \citenamefont {Sato}}]{Ito:10}%
  \BibitemOpen
  \bibfield  {author} {\bibinfo {author} {\bibfnamefont {A.}~\bibnamefont
  {Ito}}, \bibinfo {author} {\bibfnamefont {Y.}~\bibnamefont {Kozawa}}, \ and\
  \bibinfo {author} {\bibfnamefont {S.}~\bibnamefont {Sato}},\ }\href {\doibase
  10.1364/JOSAA.27.002072} {\bibfield  {journal} {\bibinfo  {journal} {J. Opt.
  Soc. Am. A}\ }\textbf {\bibinfo {volume} {27}},\ \bibinfo {pages} {2072}
  (\bibinfo {year} {2010})}\BibitemShut {NoStop}%
\bibitem [{\citenamefont {Cai}\ \emph {et~al.}(2012)\citenamefont {Cai},
  \citenamefont {Wang}, \citenamefont {Strain}, \citenamefont {Johnson-Morris},
  \citenamefont {Zhu}, \citenamefont {Sorel}, \citenamefont
  {O{\textquoteright}Brien}, \citenamefont {Thompson},\ and\ \citenamefont
  {Yu}}]{Cai2012}%
  \BibitemOpen
  \bibfield  {author} {\bibinfo {author} {\bibfnamefont {X.}~\bibnamefont
  {Cai}}, \bibinfo {author} {\bibfnamefont {J.}~\bibnamefont {Wang}}, \bibinfo
  {author} {\bibfnamefont {M.~J.}\ \bibnamefont {Strain}}, \bibinfo {author}
  {\bibfnamefont {B.}~\bibnamefont {Johnson-Morris}}, \bibinfo {author}
  {\bibfnamefont {J.}~\bibnamefont {Zhu}}, \bibinfo {author} {\bibfnamefont
  {M.}~\bibnamefont {Sorel}}, \bibinfo {author} {\bibfnamefont {J.~L.}\
  \bibnamefont {O{\textquoteright}Brien}}, \bibinfo {author} {\bibfnamefont
  {M.~G.}\ \bibnamefont {Thompson}}, \ and\ \bibinfo {author} {\bibfnamefont
  {S.}~\bibnamefont {Yu}},\ }\href {\doibase 10.1126/science.1226528}
  {\bibfield  {journal} {\bibinfo  {journal} {Science}\ }\textbf {\bibinfo
  {volume} {338}},\ \bibinfo {pages} {363} (\bibinfo {year}
  {2012})}\BibitemShut {NoStop}%
\bibitem [{\citenamefont {Huang}\ \emph {et~al.}(2020)\citenamefont {Huang},
  \citenamefont {Zhang}, \citenamefont {Xiao}, \citenamefont {Wang},
  \citenamefont {Fan}, \citenamefont {Liu}, \citenamefont {Zhang},
  \citenamefont {Qu}, \citenamefont {Ji}, \citenamefont {Han}, \citenamefont
  {Ge}, \citenamefont {Kivshar},\ and\ \citenamefont {Song}}]{Huang2020}%
  \BibitemOpen
  \bibfield  {author} {\bibinfo {author} {\bibfnamefont {C.}~\bibnamefont
  {Huang}}, \bibinfo {author} {\bibfnamefont {C.}~\bibnamefont {Zhang}},
  \bibinfo {author} {\bibfnamefont {S.}~\bibnamefont {Xiao}}, \bibinfo {author}
  {\bibfnamefont {Y.}~\bibnamefont {Wang}}, \bibinfo {author} {\bibfnamefont
  {Y.}~\bibnamefont {Fan}}, \bibinfo {author} {\bibfnamefont {Y.}~\bibnamefont
  {Liu}}, \bibinfo {author} {\bibfnamefont {N.}~\bibnamefont {Zhang}}, \bibinfo
  {author} {\bibfnamefont {G.}~\bibnamefont {Qu}}, \bibinfo {author}
  {\bibfnamefont {H.}~\bibnamefont {Ji}}, \bibinfo {author} {\bibfnamefont
  {J.}~\bibnamefont {Han}}, \bibinfo {author} {\bibfnamefont {L.}~\bibnamefont
  {Ge}}, \bibinfo {author} {\bibfnamefont {Y.}~\bibnamefont {Kivshar}}, \ and\
  \bibinfo {author} {\bibfnamefont {Q.}~\bibnamefont {Song}},\ }\href
  {http://science.sciencemag.org/content/367/6481/1018.abstract} {\bibfield
  {journal} {\bibinfo  {journal} {Science}\ }\textbf {\bibinfo {volume}
  {367}},\ \bibinfo {pages} {1018} (\bibinfo {year} {2020})}\BibitemShut
  {NoStop}%
\bibitem [{\citenamefont {Cao}\ \emph {et~al.}(2019)\citenamefont {Cao},
  \citenamefont {Chriki}, \citenamefont {Bittner}, \citenamefont {Friesem},\
  and\ \citenamefont {Davidson}}]{Cao2019}%
  \BibitemOpen
  \bibfield  {author} {\bibinfo {author} {\bibfnamefont {H.}~\bibnamefont
  {Cao}}, \bibinfo {author} {\bibfnamefont {R.}~\bibnamefont {Chriki}},
  \bibinfo {author} {\bibfnamefont {S.}~\bibnamefont {Bittner}}, \bibinfo
  {author} {\bibfnamefont {A.~A.}\ \bibnamefont {Friesem}}, \ and\ \bibinfo
  {author} {\bibfnamefont {N.}~\bibnamefont {Davidson}},\ }\href {\doibase
  10.1038/s42254-018-0010-6} {\bibfield  {journal} {\bibinfo  {journal} {Nature
  Reviews Physics}\ }\textbf {\bibinfo {volume} {1}},\ \bibinfo {pages} {156}
  (\bibinfo {year} {2019})}\BibitemShut {NoStop}%
\bibitem [{\citenamefont {Marrucci}\ \emph {et~al.}(2006)\citenamefont
  {Marrucci}, \citenamefont {Manzo},\ and\ \citenamefont
  {Paparo}}]{Marrucci2006}%
  \BibitemOpen
  \bibfield  {author} {\bibinfo {author} {\bibfnamefont {L.}~\bibnamefont
  {Marrucci}}, \bibinfo {author} {\bibfnamefont {C.}~\bibnamefont {Manzo}}, \
  and\ \bibinfo {author} {\bibfnamefont {D.}~\bibnamefont {Paparo}},\
  }\href@noop {} {\bibfield  {journal} {\bibinfo  {journal} {Phys. Rev. Lett.}\
  }\textbf {\bibinfo {volume} {96}},\ \bibinfo {pages} {163905} (\bibinfo
  {year} {2006})}\BibitemShut {NoStop}%
\bibitem [{\citenamefont {Devlin}\ \emph
  {et~al.}(2017{\natexlab{a}})\citenamefont {Devlin}, \citenamefont {Ambrosio},
  \citenamefont {Wintz}, \citenamefont {Oscurato}, \citenamefont {Zhu},
  \citenamefont {Khorasaninejad}, \citenamefont {Oh}, \citenamefont
  {Maddalena},\ and\ \citenamefont {Capasso}}]{Devlin2017Q}%
  \BibitemOpen
  \bibfield  {author} {\bibinfo {author} {\bibfnamefont {R.~C.}\ \bibnamefont
  {Devlin}}, \bibinfo {author} {\bibfnamefont {A.}~\bibnamefont {Ambrosio}},
  \bibinfo {author} {\bibfnamefont {D.}~\bibnamefont {Wintz}}, \bibinfo
  {author} {\bibfnamefont {S.~L.}\ \bibnamefont {Oscurato}}, \bibinfo {author}
  {\bibfnamefont {A.~Y.}\ \bibnamefont {Zhu}}, \bibinfo {author} {\bibfnamefont
  {M.}~\bibnamefont {Khorasaninejad}}, \bibinfo {author} {\bibfnamefont
  {J.}~\bibnamefont {Oh}}, \bibinfo {author} {\bibfnamefont {P.}~\bibnamefont
  {Maddalena}}, \ and\ \bibinfo {author} {\bibfnamefont {F.}~\bibnamefont
  {Capasso}},\ }\href@noop {} {\bibfield  {journal} {\bibinfo  {journal} {Opt.
  Express}\ }\textbf {\bibinfo {volume} {25}},\ \bibinfo {pages} {377}
  (\bibinfo {year} {2017}{\natexlab{a}})}\BibitemShut {NoStop}%
\bibitem [{\citenamefont {Devlin}\ \emph
  {et~al.}(2017{\natexlab{b}})\citenamefont {Devlin}, \citenamefont {Ambrosio},
  \citenamefont {Rubin}, \citenamefont {Mueller},\ and\ \citenamefont
  {Capasso}}]{Devlin2017J}%
  \BibitemOpen
  \bibfield  {author} {\bibinfo {author} {\bibfnamefont {R.~C.}\ \bibnamefont
  {Devlin}}, \bibinfo {author} {\bibfnamefont {A.}~\bibnamefont {Ambrosio}},
  \bibinfo {author} {\bibfnamefont {N.~A.}\ \bibnamefont {Rubin}}, \bibinfo
  {author} {\bibfnamefont {J.~P.~B.}\ \bibnamefont {Mueller}}, \ and\ \bibinfo
  {author} {\bibfnamefont {F.}~\bibnamefont {Capasso}},\ }\href@noop {}
  {\bibfield  {journal} {\bibinfo  {journal} {Science}\ }\textbf {\bibinfo
  {volume} {358}},\ \bibinfo {pages} {896} (\bibinfo {year}
  {2017}{\natexlab{b}})}\BibitemShut {NoStop}%
\bibitem [{\citenamefont {Wen}\ \emph {et~al.}(2021)\citenamefont {Wen},
  \citenamefont {Cadusch}, \citenamefont {Fang},\ and\ \citenamefont
  {Crozier}}]{Wen2021}%
  \BibitemOpen
  \bibfield  {author} {\bibinfo {author} {\bibfnamefont {D.}~\bibnamefont
  {Wen}}, \bibinfo {author} {\bibfnamefont {J.~J.}\ \bibnamefont {Cadusch}},
  \bibinfo {author} {\bibfnamefont {Z.}~\bibnamefont {Fang}}, \ and\ \bibinfo
  {author} {\bibfnamefont {K.~B.}\ \bibnamefont {Crozier}},\ }\href {\doibase
  10.1038/s41566-021-00806-x} {\bibfield  {journal} {\bibinfo  {journal}
  {Nature Photonics}\ }\textbf {\bibinfo {volume} {15}},\ \bibinfo {pages}
  {337} (\bibinfo {year} {2021})}\BibitemShut {NoStop}%
\bibitem [{\citenamefont {Nixon}\ \emph
  {et~al.}(2013{\natexlab{b}})\citenamefont {Nixon}, \citenamefont {Redding},
  \citenamefont {Friesem}, \citenamefont {Cao},\ and\ \citenamefont
  {Davidson}}]{Nixon2013a}%
  \BibitemOpen
  \bibfield  {author} {\bibinfo {author} {\bibfnamefont {M.}~\bibnamefont
  {Nixon}}, \bibinfo {author} {\bibfnamefont {B.}~\bibnamefont {Redding}},
  \bibinfo {author} {\bibfnamefont {A.~A.}\ \bibnamefont {Friesem}}, \bibinfo
  {author} {\bibfnamefont {H.}~\bibnamefont {Cao}}, \ and\ \bibinfo {author}
  {\bibfnamefont {N.}~\bibnamefont {Davidson}},\ }\href {\doibase
  10.1364/OL.38.003858} {\bibfield  {journal} {\bibinfo  {journal} {Opt.
  Lett.}\ }\textbf {\bibinfo {volume} {38}},\ \bibinfo {pages} {3858} (\bibinfo
  {year} {2013}{\natexlab{b}})}\BibitemShut {NoStop}%
\bibitem [{\citenamefont {Tradonsky}\ \emph {et~al.}(2017)\citenamefont
  {Tradonsky}, \citenamefont {Pal}, \citenamefont {Chriki}, \citenamefont
  {Davidson},\ and\ \citenamefont {Friesem}}]{Tradonsky2017}%
  \BibitemOpen
  \bibfield  {author} {\bibinfo {author} {\bibfnamefont {C.}~\bibnamefont
  {Tradonsky}}, \bibinfo {author} {\bibfnamefont {V.}~\bibnamefont {Pal}},
  \bibinfo {author} {\bibfnamefont {R.}~\bibnamefont {Chriki}}, \bibinfo
  {author} {\bibfnamefont {N.}~\bibnamefont {Davidson}}, \ and\ \bibinfo
  {author} {\bibfnamefont {A.~A.}\ \bibnamefont {Friesem}},\ }\href {\doibase
  10.1364/AO.56.00A126} {\bibfield  {journal} {\bibinfo  {journal} {Appl.
  Opt.}\ }\textbf {\bibinfo {volume} {56}},\ \bibinfo {pages} {A126} (\bibinfo
  {year} {2017})}\BibitemShut {NoStop}%
\bibitem [{\citenamefont {Pinnell}\ \emph {et~al.}(2020)\citenamefont
  {Pinnell}, \citenamefont {Nape}, \citenamefont {Sephton}, \citenamefont
  {Cox}, \citenamefont {Rodr\'{i}guez-Fajardo},\ and\ \citenamefont
  {Forbes}}]{Pinnell2020}%
  \BibitemOpen
  \bibfield  {author} {\bibinfo {author} {\bibfnamefont {J.}~\bibnamefont
  {Pinnell}}, \bibinfo {author} {\bibfnamefont {I.}~\bibnamefont {Nape}},
  \bibinfo {author} {\bibfnamefont {B.}~\bibnamefont {Sephton}}, \bibinfo
  {author} {\bibfnamefont {M.~A.}\ \bibnamefont {Cox}}, \bibinfo {author}
  {\bibfnamefont {V.}~\bibnamefont {Rodr\'{i}guez-Fajardo}}, \ and\ \bibinfo
  {author} {\bibfnamefont {A.}~\bibnamefont {Forbes}},\ }\href {\doibase
  10.1364/JOSAA.398712} {\bibfield  {journal} {\bibinfo  {journal} {J. Opt.
  Soc. Am. A}\ }\textbf {\bibinfo {volume} {37}},\ \bibinfo {pages} {C146}
  (\bibinfo {year} {2020})}\BibitemShut {NoStop}%
\bibitem [{\citenamefont {Longhi}\ and\ \citenamefont
  {Feng}(2018)}]{Longhi2018}%
  \BibitemOpen
  \bibfield  {author} {\bibinfo {author} {\bibfnamefont {S.}~\bibnamefont
  {Longhi}}\ and\ \bibinfo {author} {\bibfnamefont {L.}~\bibnamefont {Feng}},\
  }\href {\doibase 10.1063/1.5028453} {\bibfield  {journal} {\bibinfo
  {journal} {APL Photonics}\ }\textbf {\bibinfo {volume} {3}},\ \bibinfo
  {pages} {060802} (\bibinfo {year} {2018})}\BibitemShut {NoStop}%
\bibitem [{\citenamefont {Arwas}\ \emph {et~al.}(2021)\citenamefont {Arwas},
  \citenamefont {Gadasi}, \citenamefont {Gershenzon}, \citenamefont {Friesem},
  \citenamefont {Davidson},\ and\ \citenamefont {Raz}}]{arwas2021anyonic}%
  \BibitemOpen
  \bibfield  {author} {\bibinfo {author} {\bibfnamefont {G.}~\bibnamefont
  {Arwas}}, \bibinfo {author} {\bibfnamefont {S.}~\bibnamefont {Gadasi}},
  \bibinfo {author} {\bibfnamefont {I.}~\bibnamefont {Gershenzon}}, \bibinfo
  {author} {\bibfnamefont {A.}~\bibnamefont {Friesem}}, \bibinfo {author}
  {\bibfnamefont {N.}~\bibnamefont {Davidson}}, \ and\ \bibinfo {author}
  {\bibfnamefont {O.}~\bibnamefont {Raz}},\ }\href@noop {} {\bibfield
  {journal} {\bibinfo  {journal} {arXiv:2103.15359}\ } (\bibinfo {year}
  {2021})}\BibitemShut {NoStop}%
\bibitem [{\citenamefont {Mehuys}\ \emph {et~al.}(1991)\citenamefont {Mehuys},
  \citenamefont {Streifer}, \citenamefont {Waarts},\ and\ \citenamefont
  {Welch}}]{Mehuys:91}%
  \BibitemOpen
  \bibfield  {author} {\bibinfo {author} {\bibfnamefont {D.}~\bibnamefont
  {Mehuys}}, \bibinfo {author} {\bibfnamefont {W.}~\bibnamefont {Streifer}},
  \bibinfo {author} {\bibfnamefont {R.~G.}\ \bibnamefont {Waarts}}, \ and\
  \bibinfo {author} {\bibfnamefont {D.~F.}\ \bibnamefont {Welch}},\ }\href
  {\doibase 10.1364/OL.16.000823} {\bibfield  {journal} {\bibinfo  {journal}
  {Opt. Lett.}\ }\textbf {\bibinfo {volume} {16}},\ \bibinfo {pages} {823}
  (\bibinfo {year} {1991})}\BibitemShut {NoStop}%
\bibitem [{\citenamefont {Bandres}\ \emph {et~al.}(2018)\citenamefont
  {Bandres}, \citenamefont {Wittek}, \citenamefont {Harari}, \citenamefont
  {Parto}, \citenamefont {Ren}, \citenamefont {Segev}, \citenamefont
  {Christodoulides},\ and\ \citenamefont {Khajavikhan}}]{Bandres2018}%
  \BibitemOpen
  \bibfield  {author} {\bibinfo {author} {\bibfnamefont {M.~A.}\ \bibnamefont
  {Bandres}}, \bibinfo {author} {\bibfnamefont {S.}~\bibnamefont {Wittek}},
  \bibinfo {author} {\bibfnamefont {G.}~\bibnamefont {Harari}}, \bibinfo
  {author} {\bibfnamefont {M.}~\bibnamefont {Parto}}, \bibinfo {author}
  {\bibfnamefont {J.}~\bibnamefont {Ren}}, \bibinfo {author} {\bibfnamefont
  {M.}~\bibnamefont {Segev}}, \bibinfo {author} {\bibfnamefont {D.~N.}\
  \bibnamefont {Christodoulides}}, \ and\ \bibinfo {author} {\bibfnamefont
  {M.}~\bibnamefont {Khajavikhan}},\ }\href {\doibase 10.1126/science.aar4005}
  {\bibfield  {journal} {\bibinfo  {journal} {Science}\ }\textbf {\bibinfo
  {volume} {359}},\ \bibinfo {pages} {eaar4005} (\bibinfo {year}
  {2018})}\BibitemShut {NoStop}%
\bibitem [{\citenamefont {El-Ganainy}\ \emph {et~al.}(2019)\citenamefont
  {El-Ganainy}, \citenamefont {Khajavikhan}, \citenamefont {Christodoulides},\
  and\ \citenamefont {Ozdemir}}]{ElGanainy2019}%
  \BibitemOpen
  \bibfield  {author} {\bibinfo {author} {\bibfnamefont {R.}~\bibnamefont
  {El-Ganainy}}, \bibinfo {author} {\bibfnamefont {M.}~\bibnamefont
  {Khajavikhan}}, \bibinfo {author} {\bibfnamefont {D.~N.}\ \bibnamefont
  {Christodoulides}}, \ and\ \bibinfo {author} {\bibfnamefont {S.~K.}\
  \bibnamefont {Ozdemir}},\ }\href {\doibase 10.1038/s42005-019-0130-z}
  {\bibfield  {journal} {\bibinfo  {journal} {Communications Physics}\ }\textbf
  {\bibinfo {volume} {2}},\ \bibinfo {pages} {37} (\bibinfo {year}
  {2019})}\BibitemShut {NoStop}%
\bibitem [{\citenamefont {Dammann}\ \emph {et~al.}(1971)\citenamefont
  {Dammann}, \citenamefont {Groh},\ and\ \citenamefont {Kock}}]{Dammann1971}%
  \BibitemOpen
  \bibfield  {author} {\bibinfo {author} {\bibfnamefont {H.}~\bibnamefont
  {Dammann}}, \bibinfo {author} {\bibfnamefont {G.}~\bibnamefont {Groh}}, \
  and\ \bibinfo {author} {\bibfnamefont {M.}~\bibnamefont {Kock}},\ }\href
  {\doibase 10.1364/AO.10.001454} {\bibfield  {journal} {\bibinfo  {journal}
  {Appl. Opt.}\ }\textbf {\bibinfo {volume} {10}},\ \bibinfo {pages} {1454}
  (\bibinfo {year} {1971})}\BibitemShut {NoStop}%
\bibitem [{\citenamefont {Tradonsky}\ \emph {et~al.}(2019)\citenamefont
  {Tradonsky}, \citenamefont {Gershenzon}, \citenamefont {Pal}, \citenamefont
  {Chriki}, \citenamefont {Friesem}, \citenamefont {Raz},\ and\ \citenamefont
  {Davidson}}]{Tradonsky2019}%
  \BibitemOpen
  \bibfield  {author} {\bibinfo {author} {\bibfnamefont {C.}~\bibnamefont
  {Tradonsky}}, \bibinfo {author} {\bibfnamefont {I.}~\bibnamefont
  {Gershenzon}}, \bibinfo {author} {\bibfnamefont {V.}~\bibnamefont {Pal}},
  \bibinfo {author} {\bibfnamefont {R.}~\bibnamefont {Chriki}}, \bibinfo
  {author} {\bibfnamefont {A.~A.}\ \bibnamefont {Friesem}}, \bibinfo {author}
  {\bibfnamefont {O.}~\bibnamefont {Raz}}, \ and\ \bibinfo {author}
  {\bibfnamefont {N.}~\bibnamefont {Davidson}},\ }\href {\doibase
  10.1126/sciadv.aax4530} {\bibfield  {journal} {\bibinfo  {journal} {Science
  Advances}\ }\textbf {\bibinfo {volume} {5}} (\bibinfo {year} {2019}),\
  10.1126/sciadv.aax4530}\BibitemShut {NoStop}%
\bibitem [{\citenamefont {Luo}\ \emph {et~al.}(2015)\citenamefont {Luo},
  \citenamefont {Zhou}, \citenamefont {Li}, \citenamefont {Xu}, \citenamefont
  {Guo},\ and\ \citenamefont {Zhou}}]{Luo2015}%
  \BibitemOpen
  \bibfield  {author} {\bibinfo {author} {\bibfnamefont {X.-W.}\ \bibnamefont
  {Luo}}, \bibinfo {author} {\bibfnamefont {X.}~\bibnamefont {Zhou}}, \bibinfo
  {author} {\bibfnamefont {C.-F.}\ \bibnamefont {Li}}, \bibinfo {author}
  {\bibfnamefont {J.-S.}\ \bibnamefont {Xu}}, \bibinfo {author} {\bibfnamefont
  {G.-C.}\ \bibnamefont {Guo}}, \ and\ \bibinfo {author} {\bibfnamefont
  {Z.-W.}\ \bibnamefont {Zhou}},\ }\href {\doibase 10.1038/ncomms8704}
  {\bibfield  {journal} {\bibinfo  {journal} {Nature Communications}\ }\textbf
  {\bibinfo {volume} {6}},\ \bibinfo {pages} {7704} (\bibinfo {year}
  {2015})}\BibitemShut {NoStop}%
\bibitem [{\citenamefont {Perez-Garcia}\ \emph {et~al.}(2018)\citenamefont
  {Perez-Garcia}, \citenamefont {Hernandez-Aranda}, \citenamefont {Forbes},\
  and\ \citenamefont {Konrad}}]{PerezGarcia2018}%
  \BibitemOpen
  \bibfield  {author} {\bibinfo {author} {\bibfnamefont {B.}~\bibnamefont
  {Perez-Garcia}}, \bibinfo {author} {\bibfnamefont {R.~I.}\ \bibnamefont
  {Hernandez-Aranda}}, \bibinfo {author} {\bibfnamefont {A.}~\bibnamefont
  {Forbes}}, \ and\ \bibinfo {author} {\bibfnamefont {T.}~\bibnamefont
  {Konrad}},\ }\href {\doibase 10.1080/09500340.2018.1459910} {\bibfield
  {journal} {\bibinfo  {journal} {J. Mod. Opt.}\ }\textbf {\bibinfo {volume}
  {65}},\ \bibinfo {pages} {1942} (\bibinfo {year} {2018})}\BibitemShut
  {NoStop}%
\bibitem [{\citenamefont {Richardson}\ \emph {et~al.}(2013)\citenamefont
  {Richardson}, \citenamefont {Fini},\ and\ \citenamefont
  {Nelson}}]{Richardson2013}%
  \BibitemOpen
  \bibfield  {author} {\bibinfo {author} {\bibfnamefont {D.~J.}\ \bibnamefont
  {Richardson}}, \bibinfo {author} {\bibfnamefont {J.~M.}\ \bibnamefont
  {Fini}}, \ and\ \bibinfo {author} {\bibfnamefont {L.~E.}\ \bibnamefont
  {Nelson}},\ }\href {\doibase 10.1038/nphoton.2013.94} {\bibfield  {journal}
  {\bibinfo  {journal} {Nature Photonics}\ }\textbf {\bibinfo {volume} {7}},\
  \bibinfo {pages} {354} (\bibinfo {year} {2013})}\BibitemShut {NoStop}%
\bibitem [{\citenamefont {Piccardo}\ \emph {et~al.}(2021)\citenamefont
  {Piccardo}, \citenamefont {Ginis}, \citenamefont {Forbes}, \citenamefont
  {Mahler}, \citenamefont {Friesem}, \citenamefont {Davidson}, \citenamefont
  {Ren}, \citenamefont {Dorrah}, \citenamefont {Capasso}, \citenamefont
  {Dullo}, \citenamefont {Ahluwalia}, \citenamefont {Ambrosio}, \citenamefont
  {Gigan}, \citenamefont {Treps}, \citenamefont {Hiekkamäki}, \citenamefont
  {Fickler}, \citenamefont {Kues}, \citenamefont {Moss}, \citenamefont
  {Morandotti}, \citenamefont {Riemensberger}, \citenamefont {Kippenberg},
  \citenamefont {Faist}, \citenamefont {Scalari}, \citenamefont {Picqué},
  \citenamefont {Hänsch}, \citenamefont {Cerullo}, \citenamefont {Manzoni},
  \citenamefont {Lugiato}, \citenamefont {Brambilla}, \citenamefont {Columbo},
  \citenamefont {Gatti}, \citenamefont {Prati}, \citenamefont {Shiri},
  \citenamefont {Abouraddy}, \citenamefont {Alù}, \citenamefont {Galiffi},
  \citenamefont {Pendry},\ and\ \citenamefont
  {Huidobro}}]{piccardo2021roadmap}%
  \BibitemOpen
  \bibfield  {author} {\bibinfo {author} {\bibfnamefont {M.}~\bibnamefont
  {Piccardo}}, \bibinfo {author} {\bibfnamefont {V.}~\bibnamefont {Ginis}},
  \bibinfo {author} {\bibfnamefont {A.}~\bibnamefont {Forbes}}, \bibinfo
  {author} {\bibfnamefont {S.}~\bibnamefont {Mahler}}, \bibinfo {author}
  {\bibfnamefont {A.~A.}\ \bibnamefont {Friesem}}, \bibinfo {author}
  {\bibfnamefont {N.}~\bibnamefont {Davidson}}, \bibinfo {author}
  {\bibfnamefont {H.}~\bibnamefont {Ren}}, \bibinfo {author} {\bibfnamefont
  {A.~H.}\ \bibnamefont {Dorrah}}, \bibinfo {author} {\bibfnamefont
  {F.}~\bibnamefont {Capasso}}, \bibinfo {author} {\bibfnamefont {F.~T.}\
  \bibnamefont {Dullo}}, \bibinfo {author} {\bibfnamefont {B.~S.}\ \bibnamefont
  {Ahluwalia}}, \bibinfo {author} {\bibfnamefont {A.}~\bibnamefont {Ambrosio}},
  \bibinfo {author} {\bibfnamefont {S.}~\bibnamefont {Gigan}}, \bibinfo
  {author} {\bibfnamefont {N.}~\bibnamefont {Treps}}, \bibinfo {author}
  {\bibfnamefont {M.}~\bibnamefont {Hiekkamäki}}, \bibinfo {author}
  {\bibfnamefont {R.}~\bibnamefont {Fickler}}, \bibinfo {author} {\bibfnamefont
  {M.}~\bibnamefont {Kues}}, \bibinfo {author} {\bibfnamefont {D.}~\bibnamefont
  {Moss}}, \bibinfo {author} {\bibfnamefont {R.}~\bibnamefont {Morandotti}},
  \bibinfo {author} {\bibfnamefont {J.}~\bibnamefont {Riemensberger}}, \bibinfo
  {author} {\bibfnamefont {T.~J.}\ \bibnamefont {Kippenberg}}, \bibinfo
  {author} {\bibfnamefont {J.}~\bibnamefont {Faist}}, \bibinfo {author}
  {\bibfnamefont {G.}~\bibnamefont {Scalari}}, \bibinfo {author} {\bibfnamefont
  {N.}~\bibnamefont {Picqué}}, \bibinfo {author} {\bibfnamefont {T.~W.}\
  \bibnamefont {Hänsch}}, \bibinfo {author} {\bibfnamefont {G.}~\bibnamefont
  {Cerullo}}, \bibinfo {author} {\bibfnamefont {C.}~\bibnamefont {Manzoni}},
  \bibinfo {author} {\bibfnamefont {L.~A.}\ \bibnamefont {Lugiato}}, \bibinfo
  {author} {\bibfnamefont {M.}~\bibnamefont {Brambilla}}, \bibinfo {author}
  {\bibfnamefont {L.}~\bibnamefont {Columbo}}, \bibinfo {author} {\bibfnamefont
  {A.}~\bibnamefont {Gatti}}, \bibinfo {author} {\bibfnamefont
  {F.}~\bibnamefont {Prati}}, \bibinfo {author} {\bibfnamefont
  {A.}~\bibnamefont {Shiri}}, \bibinfo {author} {\bibfnamefont {A.~F.}\
  \bibnamefont {Abouraddy}}, \bibinfo {author} {\bibfnamefont {A.}~\bibnamefont
  {Alù}}, \bibinfo {author} {\bibfnamefont {E.}~\bibnamefont {Galiffi}},
  \bibinfo {author} {\bibfnamefont {J.~B.}\ \bibnamefont {Pendry}}, \ and\
  \bibinfo {author} {\bibfnamefont {P.~A.}\ \bibnamefont {Huidobro}},\
  }\href@noop {} {\bibfield  {journal} {\bibinfo  {journal} {arXiv:2104.03550}\
  } (\bibinfo {year} {2021})},\ \Eprint {http://arxiv.org/abs/2104.03550}
  {arXiv:2104.03550 [physics.optics]} \BibitemShut {NoStop}%
\bibitem [{\citenamefont {Fox}\ and\ \citenamefont {Li}(1961)}]{Fox1961}%
  \BibitemOpen
  \bibfield  {author} {\bibinfo {author} {\bibfnamefont {A.~G.}\ \bibnamefont
  {Fox}}\ and\ \bibinfo {author} {\bibfnamefont {T.}~\bibnamefont {Li}},\
  }\href {\doibase 10.1002/j.1538-7305.1961.tb01625.x} {\bibfield  {journal}
  {\bibinfo  {journal} {The Bell System Technical Journal}\ }\textbf {\bibinfo
  {volume} {40}},\ \bibinfo {pages} {453} (\bibinfo {year} {1961})}\BibitemShut
  {NoStop}%
\bibitem [{\citenamefont {Learn}\ and\ \citenamefont
  {Feigenbaum}(2016)}]{Learn2016}%
  \BibitemOpen
  \bibfield  {author} {\bibinfo {author} {\bibfnamefont {R.}~\bibnamefont
  {Learn}}\ and\ \bibinfo {author} {\bibfnamefont {E.}~\bibnamefont
  {Feigenbaum}},\ }\href {\doibase 10.1364/AO.55.004402} {\bibfield  {journal}
  {\bibinfo  {journal} {Appl. Opt.}\ }\textbf {\bibinfo {volume} {55}},\
  \bibinfo {pages} {4402} (\bibinfo {year} {2016})}\BibitemShut {NoStop}%
\bibitem [{\citenamefont {Sephton}\ \emph {et~al.}(2016)\citenamefont
  {Sephton}, \citenamefont {Dudley},\ and\ \citenamefont
  {Forbes}}]{Sephton:16}%
  \BibitemOpen
  \bibfield  {author} {\bibinfo {author} {\bibfnamefont {B.}~\bibnamefont
  {Sephton}}, \bibinfo {author} {\bibfnamefont {A.}~\bibnamefont {Dudley}}, \
  and\ \bibinfo {author} {\bibfnamefont {A.}~\bibnamefont {Forbes}},\ }\href
  {\doibase 10.1364/AO.55.007830} {\bibfield  {journal} {\bibinfo  {journal}
  {Appl. Opt.}\ }\textbf {\bibinfo {volume} {55}},\ \bibinfo {pages} {7830}
  (\bibinfo {year} {2016})}\BibitemShut {NoStop}%
\end{thebibliography}
\end{document}

% Include your paper's title here

\title{Supplementary Material to:\\
Optical vortex crystals with dynamic topologies}
%Non-Hermitian coupling in optical vortex crystals enables actively-tunable topologies
%Light vortex crystals with dynamic topologies

\author{Marco Piccardo}
\email[]{marco.piccardo@iit.it}
\affiliation{Center for Nano Science and Technology, Fondazione Istituto Italiano di Tecnologia, Milan, Italy}

\author{Michael de Oliveira}
\affiliation{Center for Nano Science and Technology, Fondazione Istituto Italiano di Tecnologia, Milan, Italy}
\affiliation{Physics Department, Politecnico di Milano, Milan, Italy}

\author{Andrea Toma}
\affiliation{Fondazione Istituto Italiano di Tecnologia, Genoa, Italy}

\author{Vincenzo Aglieri}
\affiliation{Fondazione Istituto Italiano di Tecnologia, Genoa, Italy}

\author{Andrew Forbes}
\affiliation{School of Physics, University of the Witwatersrand, South Africa}

\author{Antonio Ambrosio}
\email[]{antonio.ambrosio@iit.it}
\affiliation{Center for Nano Science and Technology, Fondazione Istituto Italiano di Tecnologia, Milan, Italy}

% Include the date command, but leave its argument blank.

% Make the title.

\maketitle 

%\tableofcontents

\section{\large{R\lowercase{oundtrip condition of the metasurface laser}}}
\label{sec_roundtripcond}
For the metasurface laser to work, the roundtrip condition has to be satisfied \cite{Wen2021}: the oscillating state has to be the same after every roundtrip inside the cavity. This means that, at steady state, the intracavity polarization and orbital angular momentum (OAM) conversions imparted by the different optical elements need to balance out, allowing to return to the same state after one cycle of transformations. In this section we show how the roundtrip condition is fulfilled in our metasurface laser by computing with Jones calculus \cite{Jones1941} the polarization and OAM state change after every optical element.

Since we work with forward and backward paths inside a cavity it is important to establish the frame of reference in which the polarization and OAM states are defined. We adopt the convention that the observer always looks at the beam following the direction of propagation \cite{Pistoni1995}, e.g. right-circularly polarized (RCP) and diagonal (D) states will become, respectively, left-circularly polarized (LCP) and anti-diagonal (A) states after reflection from a flat mirror. Similarly, the topological charge $\ell$ of an OAM state will become $-\ell$ upon reflection. Note that neither the spin angular momentum (SAM) nor OAM are changing upon reflection from the flat mirror \cite{Mansuripur2011}, the change in circular polarization state and topological charge sign are only due to the flip of the point of view of the observer.

\begin{figure*}[t]
    \centering
    \includegraphics[width=1\textwidth]{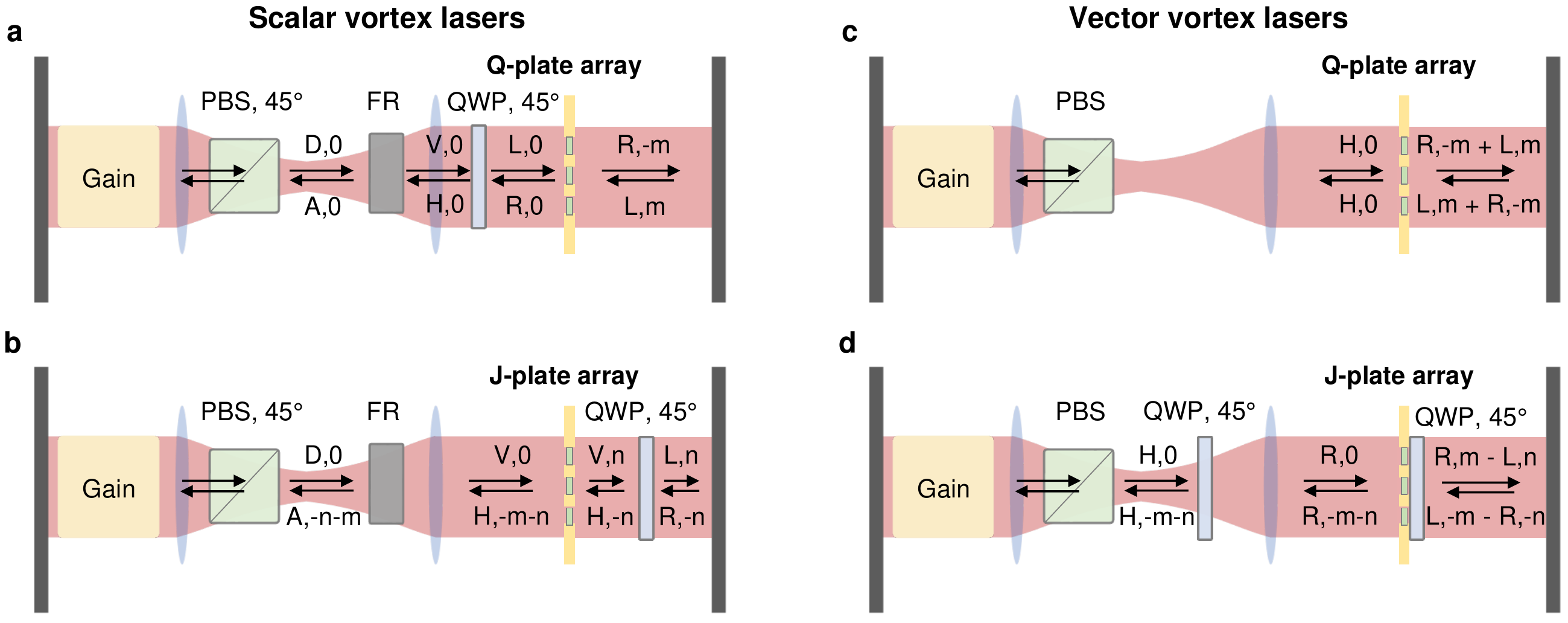}
    \caption{Roundtrip schemes of vortex lasers based on (\textbf{a}) $q$-plate, and (\textbf{b}) $J$-plate arrays, and the corresponding vector vortex laser schemes (\textbf{c},\textbf{d}), resulting in a superposition of beams with opposite spin and helicity. After every optical element the polarization state and topological charge of the OAM mode are given. PBS, polarizing beam splitter; FR, Faraday rotator; QWP, quarter waveplate; D, diagonal; A, anti-diagonal; V, vertical; H, horizontal; L, left-circular; R, right-circular.}
    \label{figS_qJroundtripschemes}
\end{figure*}

We start by illustrating different schemes producing vortex beams with uniform polarization. First, we consider an array of $q$-plates, which are reciprocal elements \cite{Wen2021} based on the Pancharatnam-Berry (or geometric) phase. The Jones matrix of a $q$-plate is
\begin{equation}
    Q = \begin{pmatrix}
        \mathrm{cos}(m\phi) & -\mathrm{sin}(m\phi) \\
        -\mathrm{sin}(m\phi) & -\mathrm{cos}(m\phi) 
        \end{pmatrix}
\end{equation}
where $m$ is the topological charge and $\phi$ the azimuthal coordinate of the beam. Since the $q$-plate operates on circular polarization states, a polarizing beam splitter (PBS) and a quarter-wave plate (QWP) are needed in the cavity to enforce circular polarization (Fig. \ref{figS_qJroundtripschemes}a). However, since the combination of PBS and QWP constitutes \textit{per se} the poor man's isolator, these elements alone would block the backward path in the cavity. To overcome this problem a Faraday rotator (FR), which is a non-reciprocal element, is added between the PBS and QWP, satisfying the roundtrip condition. To block reflections from the metallic mask encircling the metasurface array, the sample is slightly tilted by a few degrees. The output of this laser from the Talbot section of the cavity is a vortex crystal with RCP and charge $-m$. By rotating the QWP by $90^{\circ}$ the output switches to LCP with charge $m$, as verified experimentally (Fig. 2b,c of the main text). 

Next, we consider an array of $J$-plates designed to operate on a linear polarization basis \cite{Devlin2017J}. In this case the $J$-plate is purely based on the propagation phase. Its Jones matrix is
\begin{equation}
    J = \begin{pmatrix}
        e^{i m\phi} & 0 \\
        0 & e^{i n\phi} 
        \end{pmatrix}
\end{equation}
where $m$ and $n$ are two arbitrary topological charges. As shown in Fig. \ref{figS_qJroundtripschemes}b, one can use a QWP between the $J$-plate array and the mirror to switch between orthogonal linear polarization states. In this way, the reflected beam impinging on the $J$-plate will see a different topological charge, thanks to the metasurface birefringence. As shown by Jones calculus, the roundtrip condition is satisfied only if $-n-m=0$, corresponding to the relation $n = -m$ in the designed topological charges. This also shows that the laser scheme would not work for a spiral phase plate (SPP), which corresponds to a polarization independent $J$-plate with $n = m$. We will discuss a different scheme suitable for SPPs further on.

\begin{figure*}[t]
    \centering
    \includegraphics[width=1\textwidth]{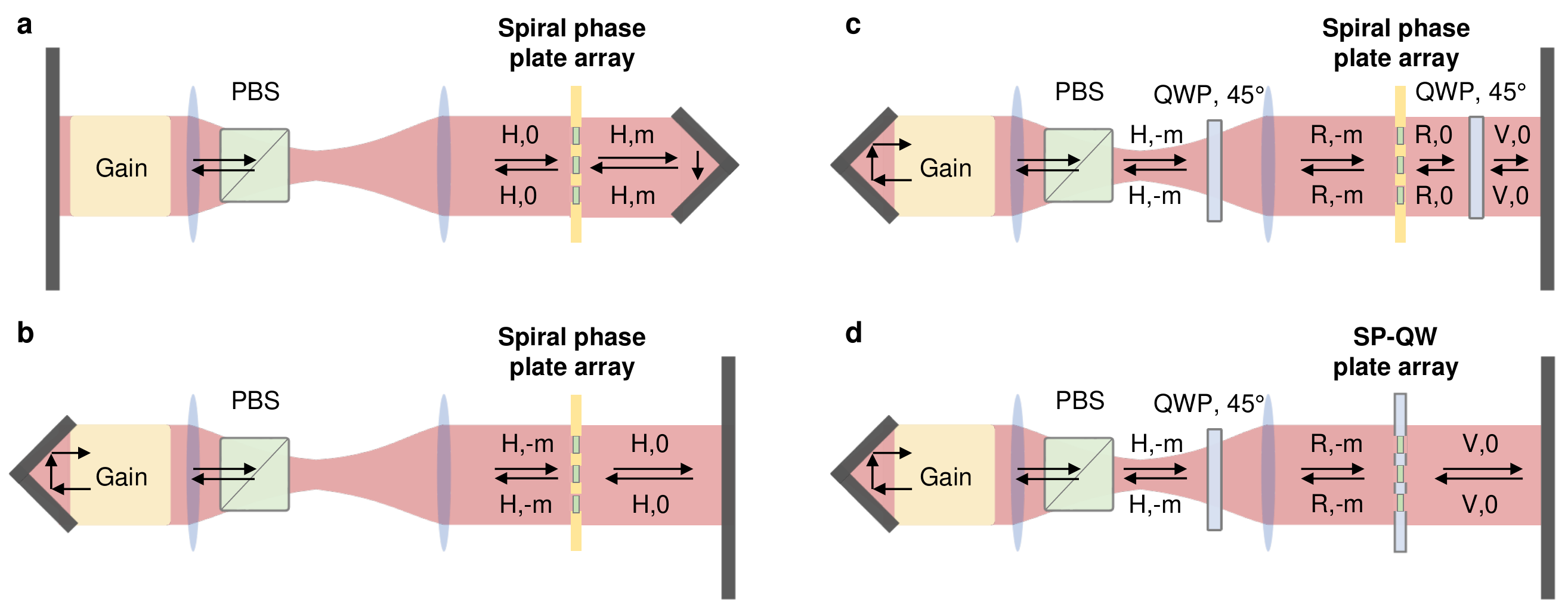}
    \caption{(\textbf{a-c}), Roundtrip schemes of vortex lasers based on spiral phase plate arrays, which require a double reflector. (\textbf{d}), Scheme based on a SP-QW metasurface integrating both a spiral phase plate and a quarter waveplate. In this case a metallic mask is not needed on the sample. After every optical element the polarization state and topological charge of the OAM mode are given. PBS, polarizing beam splitter; QWP, quarter waveplate; V, vertical; H, horizontal; R, right-circular.}
    \label{figS_SPPJQWroundtripschemes}
\end{figure*}

To complete the $q$- and $J$-plate schemes, we illustrate others that can be used to generate vector vortex beams (VVB). As shown in Fig. \ref{figS_qJroundtripschemes}c,d, the idea here is to change the polarization basis of the state impinging on the previous metasurfaces, i.e. we now use a linear polarization state for the $q$-plate and a circular polarization state for the $J$-plate. This is achieved by removing the FR and the QWP in the case of the $q$-plate, and by replacing the FR with a QWP in the case of the $J$-plate. Again, in the case of the $J$-plate the roundtrip condition is satisfied for $n=-m$. While a small tilt of the metasurface sample is needed in the schemes of Fig. \ref{figS_qJroundtripschemes}a-c, the scheme in Fig. \ref{figS_qJroundtripschemes}d allows to suppress reflections from the metallic mask without tilting the sample, thanks to the integrated poor man's isolator. The output from these lasers is a vector vortex crystal consisting of a superposition of orthogonal polarization states with different topological charge sign, i.e. an array of VVBs.

Getting back to the SPP, lasing schemes are possible by incorporating a double reflector in the cavity, such as a hollow-roof mirror or a Porro prism. The simplest cavity schemes are shown in Fig. \ref{figS_SPPJQWroundtripschemes}a,b. The key idea is that the double reflection prevents a topological charge accumulation upon a double pass through the SPP. It is important to remark that the position of the double reflector determines the side of the cavity where the vortices exist. In the configurations of Fig. \ref{figS_SPPJQWroundtripschemes}a,b, the sample needs to be slightly tilted to suppress unwanted reflections from the metallic mask. This can be avoided using the scheme of Fig. \ref{figS_SPPJQWroundtripschemes}c, which comprises a poor man's isolator, followed by a second QWP enabling the roundtrip. This second QWP could be integrated with a SPP into a single metasurface, which we call a SP-QW plate inspired by its constituting elements. Its Jones matrix is
\begin{equation}
    SPQW = \frac{e^{i m\phi}}{\sqrt{2}}\begin{pmatrix}
        1 & -i \\
        -i & 1
        \end{pmatrix}
\end{equation}
The advantage of a SP-QW is that a metallic mask is not needed on the sample, simplifying the fabrication process: in fact, only the light passing through the metasurfaces sees the embedded QWP managing to make a roundtrip. We fabricated a SP-QW and verified experimentally that the scheme in Fig. \ref{figS_SPPJQWroundtripschemes}d works. However, all the schemes in Fig. \ref{figS_SPPJQWroundtripschemes} suffer from several downsides due to the use of the double reflector. First, the double reflector does not allow to outcouple light, thus a beam splitter needs to be added inside the cavity, increasing the losses. Second, the beam reflected from the double reflector is flipped with respect to the axis of symmetry of the reflector, making the alignment with the axis of symmetry of the array critical and thus increasing the difficulty in building the cavity. Third, any double reflector  creates a loss in the reflected field at the line where the two reflecting surfaces meet \cite{Litvin2007}. In view of all these additional complications, the plane parallel resonators based on $q$- and $J$-plates shown in Fig. \ref{figS_qJroundtripschemes} are much more preferable in practice.

%\section{Lasing threshold vs. topological charge}

%The threshold of the metasurface laser is higher than that of a similar resonator implementing an array of clear apertures, without nanopillars. For instance, the threshold pump power of a metasurface laser based on a $10\times10$ $q$-plate array with $\ell=1$ is $1.7\times$ that of a laser based on an array of apertures of the same size. There are two reasons for this threshold increase. The first is due to the finite transmission of the metasurface nanopillars ($95\%$ according to our library) and diffraction losses of the nanograting. Using a telescope, we measure the transmitted power at $4f$ of a laser beam passing through the metasurface array and the aperture array, estimating the metasurface losses to $\sim40\%$. The second reason for the threshold increase originates from the self-imaging loss in the Talbot section of the cavity (Fig. \ref{figS_degeneracylift}B). By combining the Talbot and the metasurface losses, and other losses due to the intracavity optics (mainly, the $50\%$ transmission of the PBS), we obtain using an analytical roundtrip-loss model a $1.65\times$ increase in threshold pump power of the $q$-plate array with respect to the aperture array, in very good agreement with the experimental value of $1.7\times$.

%\section{Coupling strength vs. array fill factor}

\begin{figure*}[t]
    \centering
    \includegraphics[width=1\textwidth]{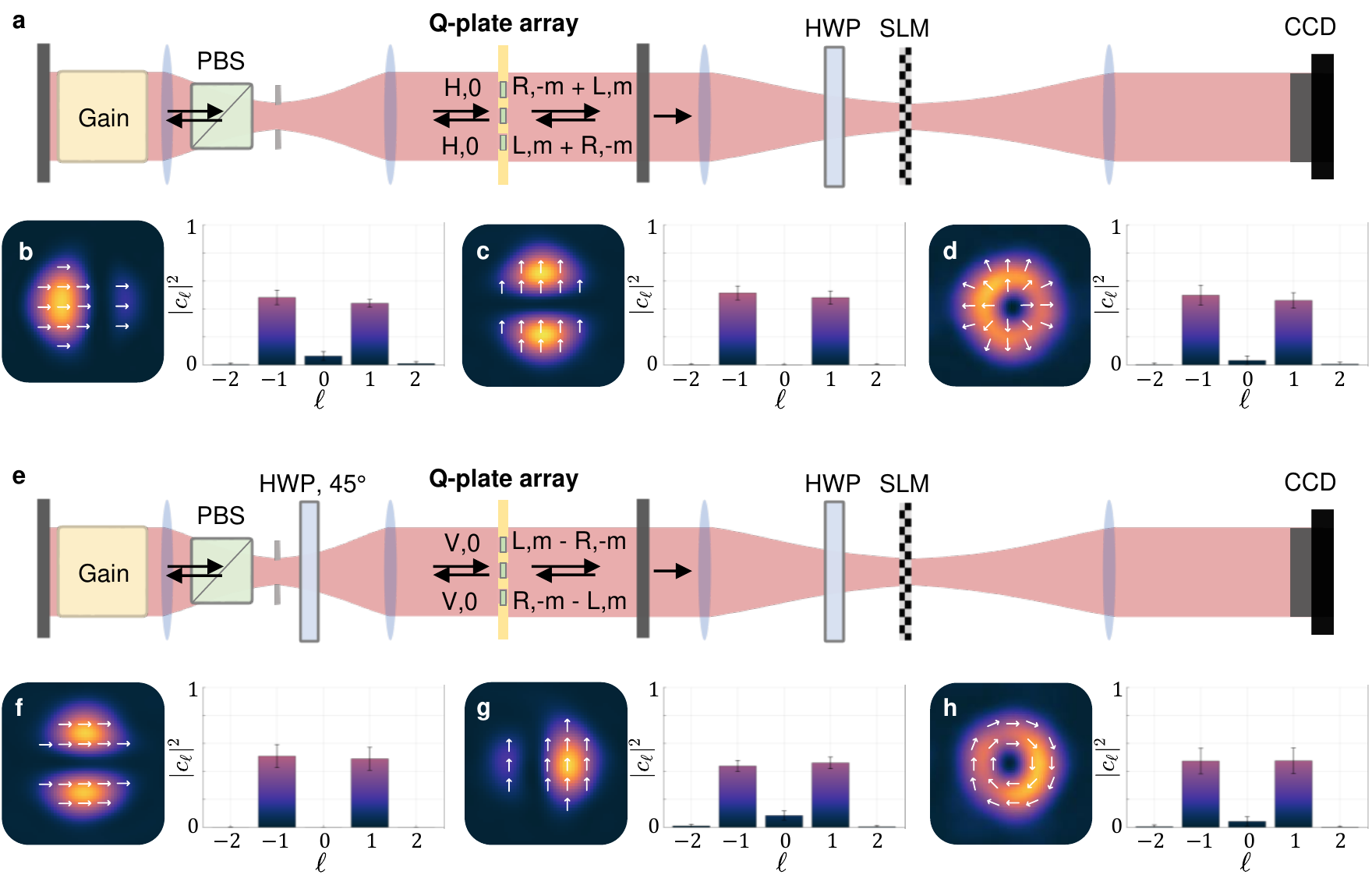}
    \caption{Topological charge analysis of azimuthally and radially polarized vector vortex crystals. (\textbf{a,e}) Schematics of the cavity in the vector vortex beam configuration. Also shown is the topological charge analysis setup. PBS, polarizing beam splitter; HWP, half-wave plate; SLM, spatial light modulator; CCD, charge-coupled device (camera); H, horizontal; V, vertical; R, right; L, left---all refer to the polarization state. $m$ is the topological charge. (\textbf{b,c,f,g}) Topological charge spectra measured by projecting the crystal in either the horizontal or vertical polarization state, as indicated by the arrows on the beam profiles. Only one beam of the array is shown, being representative of the vortex crystal. (\textbf{d,h}) Topological charge spectra of the radially and azimuthally polarized vortex crystals obtained by averaging the projective measurements for the horizontal and vertical polarization states.}
    \label{figS_azirad}
\end{figure*}

\begin{figure*}[t]
    \centering
    \includegraphics[width=1\textwidth]{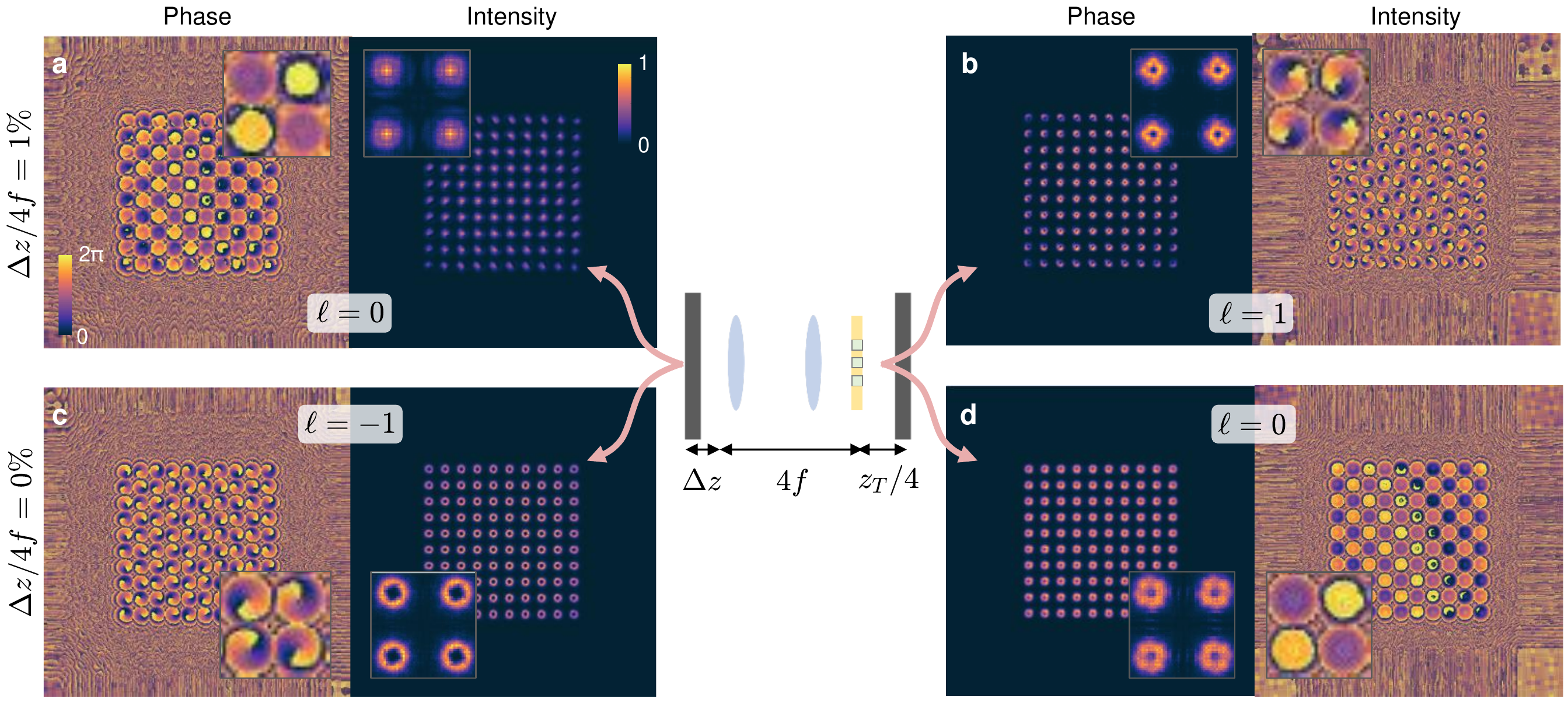}
    \caption{Phase and intensity images of crystals emitted from a metasurface laser with array charge $\ell_a=1$ as obtained from numerical Fox-Li simulations. (\textbf{a},\textbf{b}) Resonator with a $1\%$ adjustment in the length of the telescope section. (\textbf{c},\textbf{d}) Ideal self-imaging resonator without any cavity length adjustment. The topological solution changes from $(0~|~1)$ (top) to  $(-1~|~0)$ (bottom).}
    \label{figS_topologicalsolutions}
\end{figure*}

\section{\large{A\lowercase{zimuthally and radially polarized vector vortex crystals}}}

The topological charge spectra of the vector vortex crystals presented in Fig. 2d,e of the main text present contributions with zero topological charge at $\ell = 0$. These contributions originate from the finite spin-orbit conversion efficiency of the metasurfaces, which leave a fraction of the beam passing through the metasurfaces unconverted, in both polarization and OAM. When using a $q$-plate array in the scalar vortex crystal configuration (Fig. \ref{figS_qJroundtripschemes}a), this unconverted residue has a circular polarization state (either RCP or LCP) which is orthogonal to that of the vortex crystal (LCP or RCP, respectively), thus it can be polarization filtered outside the cavity, leaving no trace in the topological charge spectrum of the vortex crystal (Fig. 2b,c of the main text). On the other hand, when using a $q$-plate array in the vector vortex crystal configuration, the unconverted beam's residue has linear polarization and cannot be filtered out in the topological charge analysis.

This can be understood by examining, for instance, the characterization of radially polarized light. It is important to note that the SLM used for the charge decomposition can only modulate horizontally polarized light, due to the alignment of the constituent liquid crystals. Therefore, we carry out two charge decompositions to analyze a vector vortex beam, one per each state of the linear polarization basis. For this purpose we use a half-wave plate (HWP) before the SLM (Fig. \ref{figS_azirad}a) to project either horizontally or vertically polarized light onto the SLM (Fig. \ref{figS_azirad}b,c). As one can see, a contribution at $\ell = 0$ of around 7$\%$ (Fig. \ref{figS_azirad}b) is only present for the case of horizontal polarization. This is due to the unconverted residue of the beam impinging on the metasurface array, which is horizontally polarized and has zero OAM charge (Fig. \ref{figS_azirad}a). Analogously, in the case of azimuthally polarized light, the unconverted residue is vertically polarized (Fig. \ref{figS_azirad}e) and is present when the state is projected onto the vertical polarization basis with a value of approximately 9$\%$ (Fig. \ref{figS_azirad}g). By averaging the measurements performed in the  horizontal and vertical polarization bases, we obtain the charge spectra of the vector vortex crystals (Fig. \ref{figS_azirad}d,h).

\begin{figure*}[t]
    \centering
    \includegraphics[width=1\textwidth]{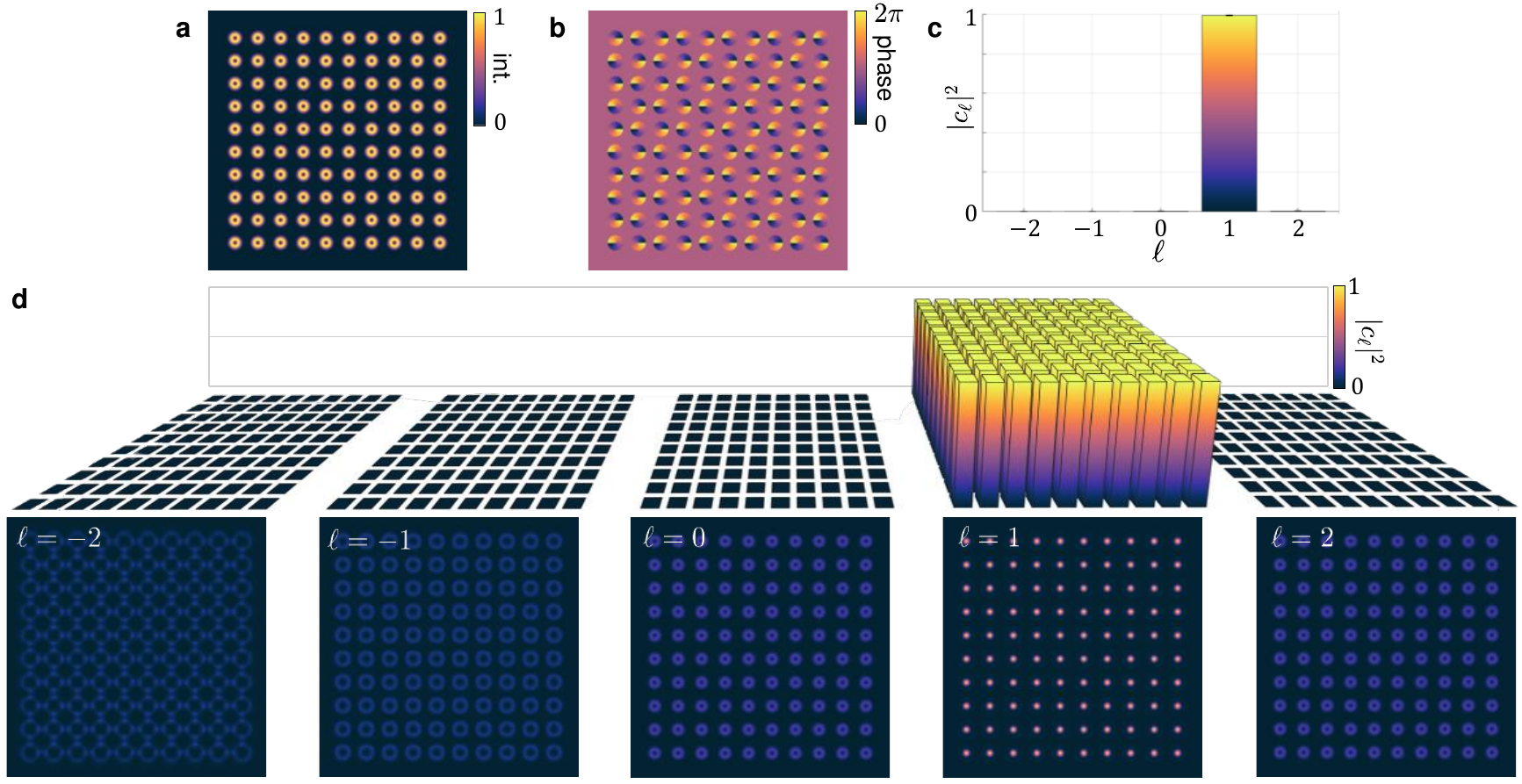}
    \caption{Parallelized topological charge characterization tested in numerical simulations. (\textbf{a}) Intensity and (\textbf{b}) phase of an array of vortices with charge 1. (\textbf{c}) Topological charge spectrum obtained from simulations replicating the experimental setup for charge characterization. (\textbf{d}) Modulated arrays obtained for different decomposition holograms.}
    \label{figS_simdecomp}
\end{figure*}

\section{\large{T\lowercase{uning the topological solutions: Laser simulations}}}

We showed in the main text, by means of experiments and calculations, that different topological solutions can exist in the metasurface laser and that these can be selected by adjusting the cavity length of the telescope with a $\Delta z$ shift with respect to the ideal $4f$ self-imaging value. The model used in the main text calculated the transmission efficiency of an array of vortices for a single roundtrip in either the telescope or Talbot section of the laser, where the metallic mask embedded in the metasurface sample acts as a spatial filter. In this section we complete the analysis by presenting the results of Fox-Li simulations, which start from a noise seed and are iterative, considering not one but many roundtrips in the laser. 

Fig. \ref{figS_topologicalsolutions} shows the phase and intensity of the crystals emitted from the metasurface laser for a relative adjustment of the cavity length of $1\%$ (A,B) and $0\%$ (C,D). Using the notation defined in the main text, in the first case the solution is $(0~|~1)$, while in the second case this becomes $(-1~|~0)$, in agreement with the experiments. It may be counter-intuitive observing donut beams in Fig. \ref{figS_topologicalsolutions}d with a topological charge of zero. This is due to the fact that the metasurface operates a phase-only transformation \cite{piccardo2020arbitrary}: the vortex beams propagating in the telescope section lose their topological charge after passing through the metasurfaces, but the intensity nulls in their beam profiles are preserved.

\section{\large{N\lowercase{umerical simulations of parallelized topological charge characterization}}}

In this work we introduce a parallelized technique that allows the topological charge spectrum of each vortex in the entire array to be analyzed all at once, regardless of the size of the array. This has the benefit of being extremely time efficient, as opposed to serial techniques analyzing one vortex at a time. Moreover, the topological spectrum is provided with spatial resolution, giving more information on the array than an average characterization analyzing the far-field with a cylindrical lens \cite{Qiao2021}. The technique is described in the Methods section and used in the main text to analyze the experimental measurements of vortex crystals. In Fig. \ref{figS_simdecomp} we apply it to a set of synthetic data obtained from numerical simulations replicating the experimental setup. We use a 10$\times$10 array of Laguerre-Gaussian modes with $p=0$ and $\ell=1$ (Fig. \ref{figS_simdecomp}a,b). The calculated topological spectrum shows that all the power is contained in the mode with charge $\ell = 1$, as expected (Fig. \ref{figS_simdecomp}c). The modulated arrays obtained for different decomposition holograms show the same features of the experiments (cf. Fig. \ref{figS_simdecomp}d and Fig. 2f of the main text), confirming the validity of the approach.

\section{\large{C\lowercase{aptions for Movies}}}

\textbf{Movie 1}---Calculated intensity (left) and phase (right) of an array of $10\times10$ vortices of topological charge $\ell=1$ propagating in the Talbot section of the cavity. $z_T$, Talbot distance.

\medskip

\textbf{Movie 2}---Calculated intensity (left) and phase (right) of an array of $10\times10$ vortices of topological charge $\ell=5$ propagating in the Talbot section of the cavity. $z_T$, Talbot distance.

\medskip

\textbf{Movie 3}---Transient showing the healing process of a topological charge defect with $\ell = 2$ (marked by a square) embedded in a metasurface array with $\ell = 1$ (right). At every roundtrip of the laser we show the dynamically-evolving phase profile of the vortex crystal. The simulation starts from a noise seed. Initially the charge corresponding to the defect device is $\ell = 2$, but after approximately 16 roundtrips the defect is healed and the charge becomes $\ell = 1$, remaining stable for any following iteration (also beyond the ones shown in the movie).

\newpage

\begin{figure*}[h!]
    \centering
    \includegraphics[width=0.8\textwidth]{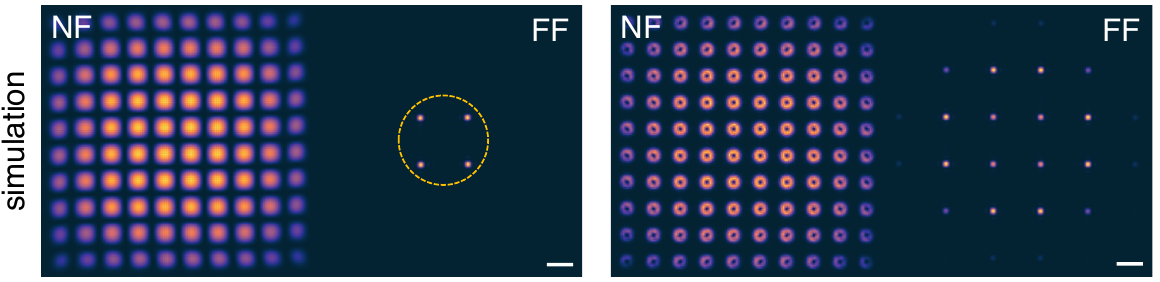}
    \caption{Simulated near-field (NF) and far-field (FF) intensity distributions of the arrays emitted from the opposite ends of the cavity, carrying a topological charge of 0 and 1. The dashed yellow line indicates the perimeter of the FF pinhole. All scale bars are 300 $\mu$m.}
    \label{figS_sim_L1}
\end{figure*}

\begin{figure*}[h!]
    \centering
    \includegraphics[width=0.5\textwidth]{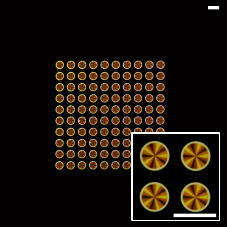}
    \caption{Optical microscope image acquired in transmission of the metasurface array with charge $\ell_a=5$ revealing the periodic azimuthal modulation. All scale bars are 300 $\mu$m.}
    \label{figS_L5microscope}
\end{figure*}